\documentclass[prd,aps,twocolumn,cite,superscriptaddress,showpacs,nofootinbib, amsmath, amssymb,preprintnumbers]{revtex4-2}

\usepackage{graphicx}
\usepackage{color}
\usepackage[utf8]{inputenc}
\usepackage[italicdiff]{physics}
\usepackage{url}

\begin{document}

\preprint{RIKEN-iTHEMS-Report-22}
\preprint{UT-HET-138}
\preprint{KANAZAWA-22-03}

\title{Constraining the primordial curvature perturbation using dark matter substructure}

\author{Shin'ichiro Ando}
\affiliation{GRAPPA Institute, Institute of Physics, University of Amsterdam, 1098 XH Amsterdam, The Netherlands}
\affiliation{Kavli Institute for the Physics and Mathematics of the Universe (WPI), University of Tokyo, Kashiwa 277-8583, Japan}

\author{Nagisa Hiroshima}
\affiliation{Department of Physics, University of Toyama, 3190 Gofuku, Toyama 930-8555, Japan}
\affiliation{RIKEN iTHEMS, Wako, Saitama 351-0198, Japan}

\author{Koji Ishiwata}
\affiliation{Institute of Theoretical Physics, Kanazawa University, Kakuma, Kanazawa, 920-1192, Japan}

\begin{abstract}

We investigate the primordial curvature perturbation by the
observation of dark matter substructure. Assuming a bump in the
spectrum of the curvature perturbation in the wavenumber of $k>1~{\rm
  Mpc}^{-1}$, we track the evolution of the host halo and subhalos in
a semianalytic way.  Taking into account possible uncertainties in
the evaluation of the tidal stripping effect on the subhalo growth, we
find a new robust bound on the curvature perturbation with a bump from
the number of observed dwarf spheroidal galaxies in our Galaxy and the
observations of the stellar stream. The upper limit on the amplitude
of the bump is $\order{10^{-7}}$ for $k\sim 10^3~{\rm Mpc}^{-1}$.
Furthermore we find the boost factor, which is crucial for the
indirect detection of dark matter signals, is up to $\order{10^4}$ due
to the bump that is allowed in the current observational bounds.

\end{abstract}
\date{\today}

\maketitle

\noindent {\it Introduction:} The observation of the cosmic microwave
background (CMB) radiation strongly supports inflation at the early
stage of the Universe. The CMB observation constrains the amplitude
$A_s$ and the spectral index $n_s$ of the scalar perturbation as
$A_s=(2.099\pm 0.029) \times 10^{-9}$ and $n_s=0.9649\pm 0.0042$ at
the pivot scale $k_*=0.05~{\rm Mpc}^{-1}$~\cite{Aghanim:2018eyx}. At a
smaller scale, on the other hand, the constraint on the curvature
perturbation is relaxed. For instance, the amplitude for the
wavenumber of $k> \order{1}~{\rm Mpc}^{-1}$ is constrained by $\mu$-
and $y$-type distortions in the CMB
observation~\cite{Chluba:2012we,Chluba:2015bqa}, the overproduction of
the primordial black holes
(PBHs)~\cite{Josan:2009qn,Carr:2009jm,Byrnes:2018txb,Dalianis:2018ymb,Sato-Polito:2019hws,Gow:2020bzo},
density profile of ultracompact
minihalos~\cite{Delos:2018ueo,Nakama:2017qac}, the free-free emission
in the Planck foreground analysis~\cite{Abe:2021mcv}, galaxy luminosity
function~\cite{Yoshiura:2020soa,Sabti:2021xvh}, and gravitational
lensing~\cite{Gilman:2021gkj}. Despite the constraints, the scalar
amplitude in the small scale can be much larger than
$\order{10^{-9}}$.  In this paper, we point out that the curvature
perturbation in such a small scale gives impact on the evolution of
the hierarchical structures of galaxies, which is traced by dark
matter halos of the Universe.

Dark matter plays a crucial role in the structure formation; the
quantum fluctuation produced by inflation seeds the density
fluctuation, which grows in the gravitational potential of dark
matter. Therefore the imprint of the small-scale perturbation during
the inflation is expected to remain in the current structure of dark
matter halos. Subhalos, which reside in larger-scale halos, are
especially promising objects to reveal the nature of dark matter.
Dwarf spheroidal galaxies (dSphs) can form inside subhalos and they
have been found and observed intensively these days in the prospects
to detect dark matter annihilation
signals~\cite{Fermi-LAT:2016uux,Hoof:2018hyn,Ando:2020yyk}.

\begin{figure}[t]
  \begin{center}
    \includegraphics[scale=0.25]{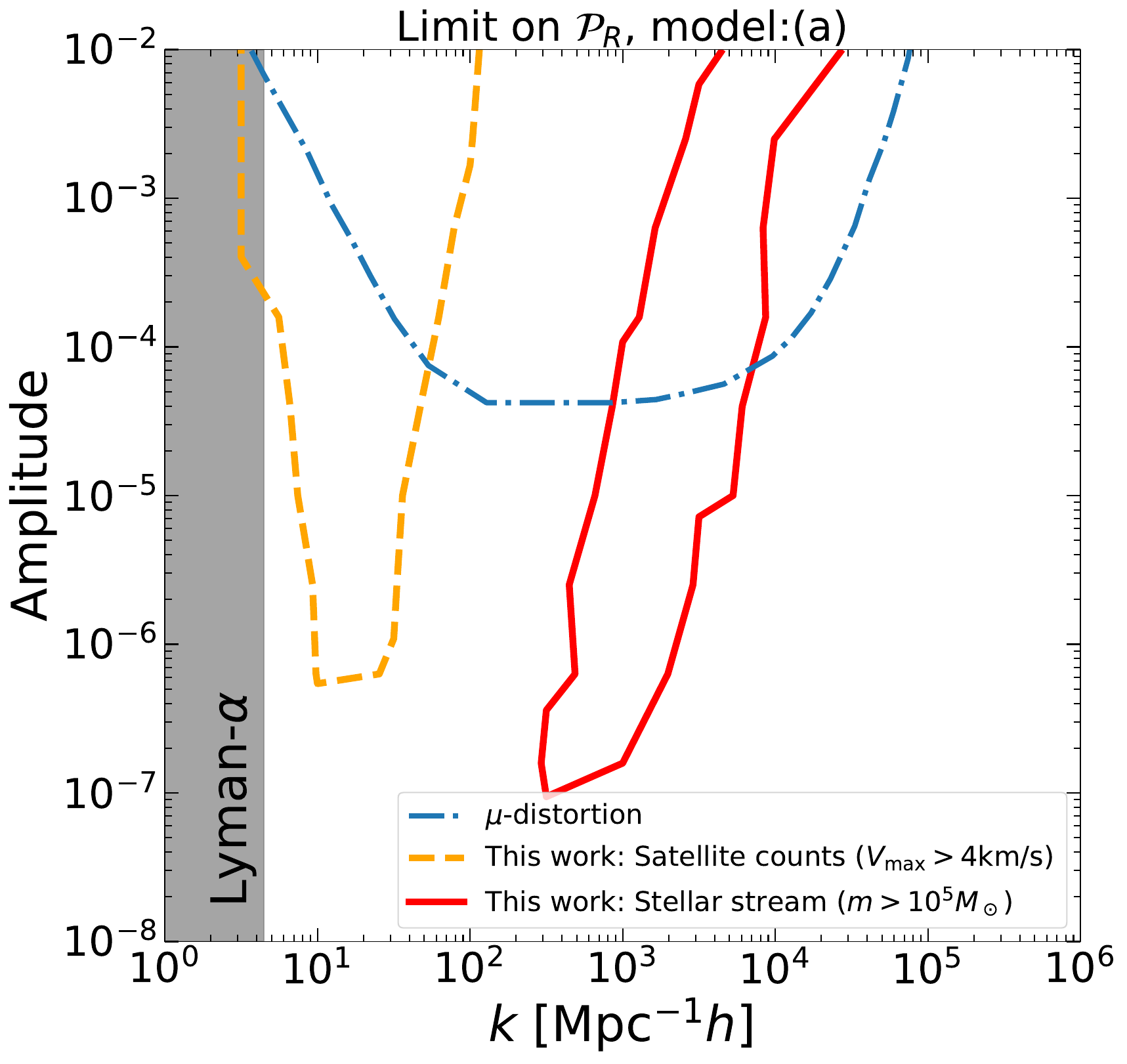}
  \end{center}
  \caption{Excluded region on the primordial curvature perturbation.
    The tidal model (a) is adopted. Upper regions separated by lines
    are excluded at 95\% C.L. ``Satellite counts'' (orange, dashed) and
    ``Stellar stream '' (red, dashed) correspond to the limits by the
    observed number of dSphs and the observation of the stellar
    stream, respectively. See Eqs.\,\eqref{eq:limit_NdSph} and
    \eqref{eq:limit_stellar}. As a reference, the constraint due to
    $\mu$-distortion is shown as ``$\mu$-distortion'', which is given
    in Ref.\,\cite{Byrnes:2018txb}. Shaded region on the left is
    disfavored from the Lyman-$\alpha$
    observations~\cite{Bird:2010mp}. }
  \label{fig:summary}
\end{figure}

In this paper we study the cosmological consequences of the primordial
curvature perturbation in the small scale. Assuming an additional bump
in the curvature perturbation, we investigate the subhalo evolution by
extending the SASHIMI
package,\footnote{\url{https://github.com/shinichiroando/sashimi-c}}
a theoretically motivated model for the tidal stripping process
calibrated by the $N$-body
simulation~\cite{Hiroshima:2018kfv,Ando:2019xlm}.  We give a new
conservative and robust bound on the curvature perturbation by using
the observed number of the dSphs in the Galactic
halo~\cite{Simon:2019nxf,Collins:2017} and the observations of the
stellar stream~\cite{Banik:2019cza,Grillmair:2006bd}. Our main result
is shown in Fig.\,\ref{fig:summary}.\footnote{See early study by
Ref.\,\cite{Dalal:2002su} which constrains the spectral index of the
scalar perturbation and neutrino masses by calculating halo evolution
and using the data of gravitational lensing.}  Additionally we give
the predictions for the annihilation boost factor, which will be
useful for future study to search for the nature of dark matter.

Throughout the paper, we adopt the cosmological parameters based on
the Planck 2018 results~\cite{Aghanim:2018eyx}
(TT,TE,EE+lowE+lensing); the density parameter of dark matter
$\Omega_{\rm dm}h^2=0.1200$ and that of baryon $\Omega_{\rm
  b}h^2=0.02237$ with $h=0.6736$.

\begin{figure*}[t]
  \begin{center}
    \includegraphics[scale=0.35]{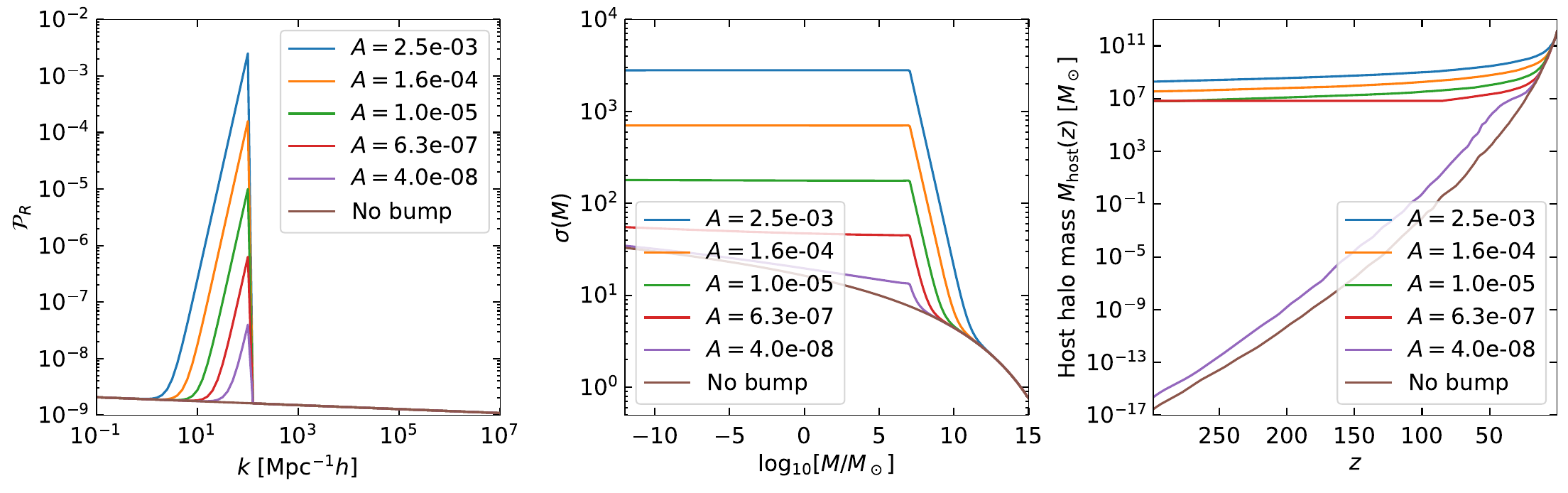}
  \end{center}
  \caption{Primordial curvature perturbation ${\cal P}_R(k)$ (left),
    variance $\sigma(M)$ of the power spectrum (middle), and average
    of host halo mass evolution $M_{\rm host}(z)$
    (right). $k_b=1.0\times 10^{2}~{\rm Mpc}^{-1}h$ is taken for all
    and each line corresponds to $A=2.5\times 10^{-3}$, $1.6\times
    10^{-4}$, $1.0\times 10^{-5}$, $6.3\times 10^{-7}$, and $4.0\times
    10^{-8}$. As a reference, the result without the bump is shown as
    `No bump'. $M_{\rm host}(0)=1.3\times10^{12}M_\odot$ is taken for
    $M_{\rm host}(z)$. See Appendix~\ref{app:figs} for additional
    figures with different values of $k_b$.}
  \label{fig:1}
\end{figure*}

\noindent {\it The primordial curvature perturbation:} We consider a
model in which the primordial power spectrum has a bump in the small
scale $k\gtrsim \order{1}~{\rm Mpc}^{-1}$.
In order to investigate the impact of the curvature perturbation in
the region, we consider an additional bump on top of the nearly
scale-invariant curvature perturbation that is consistent with the CMB
observation:
\begin{align}
  {\cal P}_R= {\cal P}_{R}^{(0)} + {\cal P}_{R}^{\rm bump}\,,
\end{align}
where ${\cal P}_{R}^{(0)}(k) = A_s (k/k_*)^{n_s-1}$ and
\begin{align}
  {\cal P}_{R}^{\rm bump}(k;k_b) =
    \left\{
    \begin{array}{ll}
      (A-{\cal P}_{R}^{(0)}(k_b))
      \left(\frac{k}{k_b}\right)^{n_b} & k \le k_b \\
    0 & k>k_b
    \end{array}
    \right.\,.
\end{align}
Here we have introduced three parameters, $A$, $k_b$, and $n_b$. In
Ref.\,\cite{Byrnes:2018txb} the steepest spectral index is $n_b=4$ in
single-field inflation. On the other hand, Ref.\,\cite{Ozsoy:2019lyy}
claims that the spectral index can be as large as $8$ after
encountering a dip in the amplitude and then the amplitude reaches to
a peak with the index less than 4. In our study we adopt $n_b=4$ and
take $A$ and $k_b$ as free parameters. We plot several examples of the
${\cal P}_R$ in Fig.\,\ref{fig:1}. The parameters are given in the
figure caption.

\begin{figure}[t]
  \begin{center}
    \includegraphics[scale=0.35]{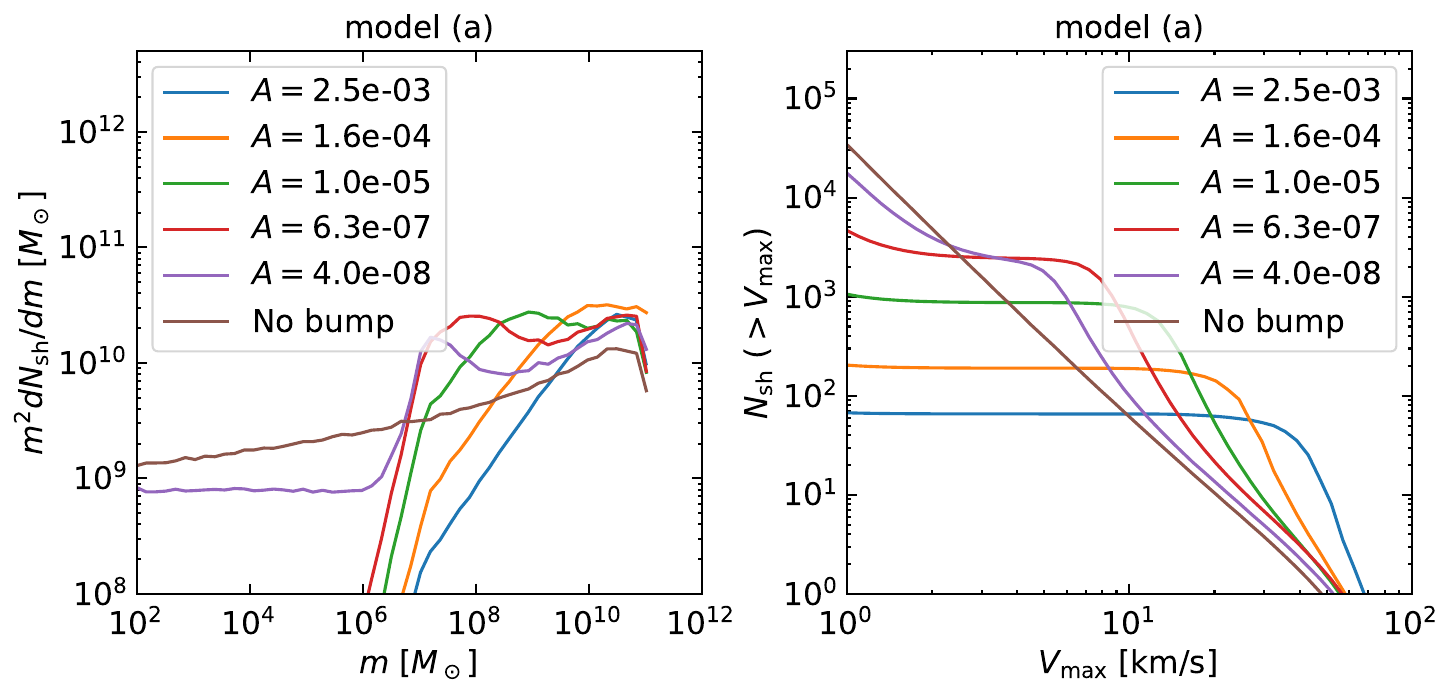}
  \end{center}
  \caption{Mass function $dN_{\rm sh}/dm$ of subhalo (left) and
    cumulative maximum circular velocity function $N_{\rm sh} (>V_{\rm
      max})$ (right). We take tidal model (a) and the other parameters
    are the same as Fig.\,\ref{fig:1}. See Appendix~\ref{app:figs} for
    additional figures with different values of $k_b$.}
  \label{fig:2}
\end{figure}

From the curvature perturbation, the variance of the linear power
spectrum in the comoving scale $R$ is given by
\begin{align}
  \sigma^2(M) = \int d\ln k~\frac{k^3}{2\pi^2} P(k) W^2(kR)\,,
\end{align}
where $P(k)$ is the power spectrum calculated from ${\cal P}_R$ and
$W(kR)$ is the window function. We adopt the sharp-$k$ window,
$W(kR)=\Theta(1-kR)$, where $\Theta$ is the Heaviside step
function. This is because for the power spectrum that has a steep
cutoff, it is shown in Ref.\,\cite{Schneider:2013ria} that the
sharp-$k$ window gives a good agreement with the
simulation.\footnote{We have calculated the variance by using the
top-hat window.  The variance becomes a smoother function and the halo
evolution merely changes. Thus the exclusion limits, which we see
later, do not change.}  The mass scale~$M$ is given as
$M=(4\pi/3)(Rc)^3 \rho_m$ where $\rho_m=\Omega_m \rho_c$ ($\rho_c$ is
the critical density) and a parameter $c=2.7$ is determined by
comparing with the simulation~\cite{Schneider:2013ria}.

The result of $\sigma(M)$ is given in Fig.\,\ref{fig:1}.  It is seen
that the bump with $A\gtrsim 10^{-6}$ significantly affects the
variance.\footnote{A similar variance is obtained due to the formation
of the PBHs~\cite{Kadota:2020ahr}.}  $\sigma$ is enhanced below a mass
scale, which is, for instance, $(4\pi/3)(c/k_b)^3 \rho_m\sim
10^{7}\,M_\odot$ for $k_b=10^2\,{\rm Mpc}^{-1}h$. The scale gets
smaller as $k_b$ becomes larger. The variance becomes almost constant
for $A\gtrsim 10^{-6}$. This is because the bump contributes
dominantly in the integral below the scale $k_b$.

\noindent {\it The host halo and subhalo evolution:} The enhancement
on the variance of the power spectrum due to the bump can affect the
merger history of both the host and subhalos.  To see this we evaluate
the halo evolution history based on the extended Press-Schechter (EPS)
formalism~\cite{Lacey:1993iv,Yang_2011}. See Appendixes~\ref{app:EPS}
and \ref{app:subhalo} for details.

The host halo mass evolution $M_{\rm host}(z)$ is shown in
Fig.\,\ref{fig:1}, where the average of 200 host halo realizations is
given.  We have checked that the result without the bump agrees with
the fitting formula given in Ref.\,\cite{Correa:2014xma} in the low
$z$ region, which is calibrated against the simulations in
$z\lesssim10$.  It is seen that at the low $z$ region the host halo
evolution coincides with the one without the bump. However, the
difference becomes significant in high $z$ region, especially for a
large $A$; we find a plateau for large values of $A$, such as
$A\gtrsim 10^{-6}$.  In the conventional model, i.e., with no bump in
the curvature perturbation, the halos with small mass scales are
formed in the past and they grow to massive halos to the
present. This is true for the model with bump with a small amplitude,
such as $A\lesssim 10^{-8}$.  When the amplitude is large, on the
other hand, the halo formation and growth happen almost at a certain
redshift. For $k_b=10^{2}\,{\rm Mpc}^{-1}h$, for instance, halos with
mass $\sim 10^7\,M_\odot$ form at once. We see a mild dependence of
the amplitude on the mass scale; the mass scale is larger for the
larger amplitude.  This is because $\sigma$ is altered in the larger
mass-scale region as the amplitude is larger. After the formation they
merely grow for some period since $\sigma$ changes drastically for the
small-mass scales. This period corresponds to the plateau for $M_{\rm
  host}$. Eventually, the host mass rebegins to grow in accordance
with the case without the bump, which is a reasonable behavior since
$\sigma$ coincides with the one calculated without the bump.

The evolution of the subhalos is similar to that of the host halo, but
it suffers from the tidal stripping due to the gravitational potential
of the host halo after the accretion.  To evaluate the subhalo
evolution, we modify SASHIMI package to implement the results of
$\sigma(M)$ and the host halo evolution $M_{\rm host}(z)$ with the
existence of a bump in the primordial curvature perturbation. For
consistency, we adopt the concentration-mass relation given by
Ref.~\cite{Sanchez-Conde:2013yxa}, where the fitting function is given
in terms of $\sigma(M)$. In the code the Navarro-Frenk-White (NFW)
profile~\cite{Navarro:1995iw} with truncation is assumed, which is
characterized by typical mass density $\rho_{\rm s}$, the scale radius
$r_{\rm s}$ and the truncation radius $r_t$. For the tidal process, we
consider three models;
\begin{itemize}
\item[(a)] $g=0.86$ and $\zeta=0.07$~\cite{Jiang:2014nsa},
\item[(b)] $z$-dependent $g(z)$ and $\zeta(z)$~\cite{Hiroshima:2018kfv},
\item[(c)] no tidal stripping,
\end{itemize}
where $g$ and $\zeta$ are the parameters in the evolution of the
subhalo mass~\cite{Jiang:2014nsa}
\begin{align}
  \dv{m}{t} = -g \frac{m}{\tau_{\rm dyn}}\left(\frac{m}{M(z)}\right)^\zeta\,.
\end{align}
Here $\tau_{\rm dyn}$ is the halo's dynamical time. Since $g(z)$ and
$\zeta(z)$ are given in $z\le 7$ in model~(b), we take $g(z)=g(7)$ and
$\zeta(z)=\zeta(7)$ for $z>7$ in the current calculation. The
model~(c) corresponds to the so-called unevolved mass function.  We
note that the model (c) might be more realistic than the others when
we compute the boost factor. This is because the tidal stripping
effect may not change the inner structure of the halo profile in the
case of a highly concentrated profile~\cite{Delos:2019lik}, which is
expected in the current case. The mass distribution
function of subhalos at the accretion is given by the EPS
formalism as a function of $m_a$, $z_a$ and the host halo mass $M_0$ at
$z=z_0$.  Using the number $d^2N_{{\rm sh},a}$ of subhalos with mass
$m_a$ that accrete at $z=z_a$, the subhalo mass function after the
tidal stripping is obtained by
\begin{align}
  \frac{dN_{{\rm sh}}}{dm}&=
  \int d^2N_{{\rm sh},a}
  \int dc_{{\rm vir},a} P^{c_{{\rm vir},a}}(m_a,z_{a})\delta(m-m_0)\,,
\end{align}
where $P^{c_{{\rm vir},a}}(m_a,z_{a})$ is the distribution function
for $c_{{\rm vir},a}$ that is computed from the one for the
concentration-mass relation. A subscript ``0'' stands for the values
at $z=z_0$.

We plot the mass function $dN_{\rm sh}/dm $ of the subhalo at $z_0=0$
computed using the tidal model (a) in Fig.\,\ref{fig:2}.  We found
that the mass function is affected significantly, depending on $A$ and
$k_b$.  As $k_b$ becomes small, the mass function is altered in the large
subhalo mass, which is expected from the behavior of $\sigma(M)$. Due
to the bump, the number of the subhalo of a mass scale tends to be
enhanced.  On the contrary, the mass function is suppressed below that
mass scale. This effect is significant for large $A$ and small
$k_b$. Such a drastic change leads to change the prediction of the
number of dSphs, which are formed in subhalos. Additionally, we found
that the result is almost independent of the tidal models and $z_{\rm
  max}$ if $z_{\rm max}\ge 7$. Here $z_{\rm max}$ is the maximum
redshift to track the subhalo evolution. Therefore, we expect the
observable consequences are determined by the evolution in the low
redshift regime and that they are not significantly affected by the
details of the tidal evolution models, which we confirm below.

\noindent {\it Astrophysical observables and constraint:} It is
considered that subhalos which satisfy certain conditions form
galaxies inside. A quantity for the criterion is the maximum circular
velocity.  Based on the conventional theory of galaxy formation, for
instance, the dSphs formation occurs for $V_{{\rm max},a}>18~{\rm
  km/s}$, where $V_{{\rm max},a}$ is the maximum circular velocity at
the time of the accretion. Using this condition, we can predict the
number of the present dSphs in the Galaxy, which is one of the
important observables of the dSphs.  However, this criterion is under
debate.  A recent study suggests a different criterion of $V_{\rm
  max,a}>10.5~{\rm km/s}$~\cite{Graus_2019}. Therefore, the predicted
number of dSphs can change, depending on the choice of criteria.  To
avoid such uncertainties on the condition for formation of a dSph, we
take a much conservative approach based on the observations. Focusing
on the present maximum circular velocity $V_{\rm max}$ that is
actually observed for dSphs, the number of dSphs in our Galaxy whose
$V_{\rm max}$ is over 4 km/s is given
by~\cite{Dekker:2021scf,DES:2019vzn}
\begin{align}
  N_{\rm dSph}^{\rm low}(V_{\rm max}>4~{\rm km/s})=94\,.
\end{align}
The condition $V_{\rm max}>4\ {\rm km/s}$ is determined by the minimum
of the observed velocity dispersions among
dSphs~\cite{Simon:2019nxf,Collins:2017}. We apply this observational
bound directly to our calculation by imposing the following
condition
\begin{align}
  N_{\rm sh}(V_{\rm max}>4~{\rm km/s})\ge
  N_{\rm dSph}^{\rm low}(V_{\rm max}>4~{\rm km/s})\,,
  \label{eq:limit_NdSph}
\end{align}
and see whether the curvature perturbation with the bump contradicts
the observation. Following the method adopted in
Ref.\,\cite{Dekker:2021scf}, we derive 95\% C.L. exclusion limit.

Fig.\,\ref{fig:2} shows the cumulative maximum velocity function
$N_{\rm sh}\,(>V_{\rm max})$ of subhalos.  It is seen that $N_{\rm
  sh}\,(>V_{\rm max})$ is enhanced compared to the case with no bump
at a large value of $V_{\rm max}$, corresponding to a massive case, as
$A$ becomes large and $k_b\lesssim 10^2\,{\rm Mpc}^{-1}h$. In the
exchange for the enhancement at a large $V_{\rm max}$ region, it is
suppressed in small $V_{\rm max}$ region. For instance, it is below
the observed value for $k_b=10^2\,{\rm Mpc}^{-1}h$ and $A\gtrsim
2.5\times 10^{-3}$. Therefore, the bump in the primordial curvature
perturbation in such a parameter space is excluded. On the other hand,
for $k_b\gtrsim 10^2\,{\rm Mpc}^{-1}h$, the cumulative maximum
circular velocity function is almost unchanged. This reflects the fact
that the bump with a large $k_b$ affects less massive halos than those
responsible for dSphs.

Making a comprehensive analysis on the bump model, we compute the
cumulative number of the subhalos $N_{\rm sh}\,(V_{\rm max}>4~{\rm
  km/s})$ on the $(A,\,k_b)$ plane.  We found that it is smaller than the
observed value in the region $A\gtrsim 10^{-6}$ and $k_b\lesssim
10^2\,{\rm Mpc}^{-1}h$ so that the region is excluded. The result is
shown in Fig.\,\ref{fig:summary}. As expected from the results of the
mass function, we confirmed that the number of subhalos satisfying the
condition of the maximum circular velocity and the resultant exclusion
region do not depend on the tidal stripping models. Therefore, it is
concluded that the bound is conservative and robust.

Another observable effect appears in the stellar stream, where gaps
are caused by a passage of subhalos in the Galaxy. A too large or too
small number of subhalos may conflict with the observation of the
stellar stream.  We make use of the results by
Ref.\,\cite{Banik:2019cza}, which analyzes the GD-1
stream~\cite{Grillmair:2006bd} using data from
Gaia~\cite{Gaia:2016,Gaia:2018} and Pan-STARRS
survey~\cite{Pan-STARRA1:2016}. We adopt the most conservative limit
on the number of subhalos whose mass is within
$10^5M_\odot$\,--\,$10^9M_\odot$, which is given by
\begin{align}
  N_{\rm sh}/N_{\rm sh, CDM}<2.7~(95\%~{\rm C.L.})\,,
  \label{eq:limit_stellar}
\end{align}
where $N_{\rm sh, CDM}$ corresponds to the one without the
bump. Consequently we found that the amplitude in $k_b=
\order{10^2\,\mathchar`-\,10^{4}}\,{\rm Mpc}^{-1}h$ is constrained; the
most stringent upper limit is $10^{-7}$, which is shown in
Fig.\,\ref{fig:summary}.

The limits from the observations of satellite number and stellar stream for
tidal models (b) and (c) are given in Appendix~\ref{app:figs}. We found that
the bounds are almost unchanged by choice of the tidal models. Therefore,
the limits shown in Fig.\,\ref{fig:summary} are conservative and robust. 

Finally we discuss the annihilation boost factor due to the
substructure in the host halo. The enhancement of the subhalo clustering
leads to a large enhancement of the pair-annihilation signals of dark
matter.  We define the boost factor as $ B\equiv J_{{\rm sh}}^{\rm
  tot}/J_{\rm h}$, where
\begin{align}
  J_{\rm h}=\int d^3x \rho_{\rm h}^2\,,~~
  J_{\rm sh}^{\rm tot}&= \int d\rho_{\rm s} dr_{\rm s} dr_t
  \frac{d^3N_{\rm sh}}{d\rho_{\rm s} dr_{\rm s} dr_t}
  \int d^3x  \rho_{\rm sh}^2\,,  
\end{align}
assuming the NFW profile $\rho_{\rm h}$ and $\rho_{\rm sh}$ for both
the host and subhalos, respectively.

Note that the boost factor depends on the minimum halo
mass~\cite{Sanchez-Conde:2013yxa}.  In our analysis we take minimum
halo mass as $10^{-6}M_\odot$, assuming neutralino-like dark
matter~\cite{Diemand:2005vz}.  We found that the boost factor is
significantly enhanced by the amplitude of the bump in the regions
which are not excluded from current observations. (See
Appendix~\ref{app:figs} for details.) It becomes as large as
$\order{10^4}$ for $A\sim \order{10^{-2}}$ and $k_b\sim
\order{10^{5}}$ Mpc$^{-1}h$. Additionally, we observe a mild
dependence on $k_b$, i.e, the boost factor gets larger for larger
$k_b$.

Careful readers may think that the resultant boost factor in
the model (c) should be enhanced compared to the model (a) or (b)
since the tidal stripping process reduces the subhalo mass. Although
the subhalos lose their masses due to the tidal process, the inner
structure of the subhalo is hardly affected. This is because the
concentration parameter is much larger than
$\order{1}$~\cite{Sanchez-Conde:2013yxa}. The concentration-mass
relation at high redshifts is still under debate (e.g., see
Ref.\,~\cite{Wang:2022spb}). Hence the model of the concentration-mass
relation would be the main source of the uncertainty in the estimation
of the boost factor.  If the issue is settled, then the clustering of
subhalos would be another important observable to constrain the
unconventional curvature perturbation in the future experiment.

\noindent {\it Conclusion:} In this work, we propose a new scheme for
investigating the primordial curvature perturbation of the small scale
by using the observation of the dark matter substructure.  Assuming an
additional bump in the primordial curvature perturbation, an
enhancement on the variance of the linear power spectrum appears. We
track the evolution of both the host halo and subhalos from the power
spectrum, based on the EPS formalism and the semianalytic
calculation. In the evolution of the subhalo, we take into account
uncertainty in the evaluation of the tidal stripping effect by
comparing three types of models. We have found that the extra bump in
the curvature perturbation significantly affects the evolution of the
host and subhalos, depending on the amplitude and the wavenumber scale
of the bump. On the other hand, it turns out that the mass functions
of the subhalos merely depend on the tidal models. This fact enables
us to compute the number of dSphs that is free from the uncertainty in
the tidal models. We focus on the cumulative number of subhalos with
maximum circular velocity over $4~{\rm km/s}$, which is observed
directly.  Imposing the most conservative bound from observations of
the dSphs, we have found that the bump with amplitude $A\gtrsim
10^{-6}$ in $k\lesssim 10^2~{\rm Mpc}^{-1}h$ is excluded.  Another
consequence that has a direct connection with the number of subhalos
is the stellar stream. Adopting the most conservative limit from the
observation, the amplitude of the bump in the region
$k=\order{10^2\,\mathchar`-\,10^{4}}\,{\rm Mpc}^{-1}h$ is constrained
as $A\gtrsim 10^{-7}$ at most. We also obtain an indication for the
boost factor, which is crucially important for detecting dark matter
annihilation signals. The boost factor of $\order{10^4}$ can be
expected in the parameter region allowed by the existing
observations. In the future, the predictions in this work could be
tested by various probes of small-scale halos, such as gravitational
lensing
observations~\cite{Vegetti:2018dly,Gilman:2019vca,Gilman:2019nap,Nadler:2021dft,Montel:2022fhv}
or pulsar timing array
experiments~\cite{Lee:2020wfn,Lee:2021zqw,Kashiyama:2018gsh,Delos:2021rqs,Clark:2015sha,Ishiyama:2010es}. We
leave it for future work.

\section*{acknowledgment}
We thank T.~Ishiyama, T.~Sekiguchi and K.~Yang for valuable
discussion.  This work is supported by JSPS KAKENHI Grants
No. JP19K23446 and No. JP22K14035 (N. H.), MEXT KAKENHI Grants
No. JP20H05852 (N.H.), No. JP20H05850, and No. JP20H05861 (S.A.). The work
of K.I. is supported by JSPS KAKENHI Grants No. JP18H05542, No. JP20H01894,
and JSPS Core-to-Core Program Grant No. JPJSCCA20200002. This work was
supported by computational resources provided by iTHEMS.

\begin{widetext}

\appendix

\section{The extended Press-Schechter formalism}
\label{app:EPS}

The collapse of the overdensity to form the halos is characterized by
two quantities:
\begin{align}
  \delta(z) = \delta_c/D(z)\,,~~~
  \sigma(M)\,,
\end{align}
where $z$ is the redshift, $\delta_c\simeq 1.686$ is the critical
overdensity, and $D(z)$ is the linear
growth factor defined by
\begin{align}
  D(z) = D_{\rm norm} H(z) \int_z^\infty dz'\,\frac{1+z'}{H^3(z')}\,.
\end{align}
Here $D_{\rm norm}$ is determined to satisfy $D(0)=1$ and $H(z)$ is
the Hubble parameter. Since $D(z)$ is monotonically decreasing
function of $z$, $\delta(z)$ monotonically decreases as $z$ gets
small. On the other hand, $\sigma(M)$ is cumulative as $M$ becomes
small. Namely, the parameters $z$ and $M$ can be translated into
$\delta(z)$ and $\sigma(M)$, respectively.  Based on the EPS theory,
the evolution of the halo is described by the following probability
distribution function
(PDF)~\cite{Bond:1990iw,Bower:1991kf,Lacey:1993iv},
\begin{equation}
    f(S_2,\delta_2|S_1,\delta_1)=
    \frac{1}{\sqrt{2\pi}}\frac{\delta_2-\delta_1}{\left(S_2-S_1\right)^{3/2}}\exp\left[-\frac{\left(\delta_2-\delta_1\right)^2}{2\left(S_2-S_1\right)}\right]\,,
    \label{eq:f}
\end{equation}
where $S_i\equiv\sigma^2(M_i)$ and $\delta_i\equiv\delta(z_i)$ and
$z_1<z_2$. Namely, $f(S_2,\delta_2|S_1,\delta_1)$ is the PDF where a
halo with a mass of $M_2$ is a progenitor at $z=z_2$ of a halo with a
mass of $M_1$ at $z=z_1$. In the canonical $\Lambda$CDM case, halos
with small mass are created in the past and evolve by accretions and
mergers to the present.  With the additional bump in the primordial
curvature perturbation, however, this picture changes.

In the calculation, we construct the evolution history of the host
halo by applying the inverse function method to the above distribution
function (see details for Ref.~\cite{Hiroshima:2022khy}).  We start
our calculation of the host halo evolution from $z=0$ and track its
merger history up to $z=300$, taking 4000 points of a constant
interval in the $\ln(1+z)$ space. Considering the Milky Way-like host
halo at $z=0$, we take the host halo mass as $M_{\rm
  host}(z=0)=1.3\times10^{12}M_\odot$~\cite{Bland-Hawthorn:2016}.

\section{Subhalo evolution}
\label{app:subhalo}

To begin with, we collect important quantities for the subhalo
properties. We assume that the halos follow the Navarro-Frenk-White
(NFW) profile~\cite{Navarro:1995iw} with truncation, which is
characterized by typical mass density $\rho_{\rm s}$, the scale radius
$r_{\rm s}$ and the truncation radius $r_t$ as
\begin{align}
  \rho_{\rm NFW}(r) = 
  \left\{
  \mqty{
  \frac{\rho_{\rm s}}{(r/r_{\rm s})(1+r/r_{\rm s})^2} &~~~~ r\le r_t \\
  0 &~~~~ r>r_t }
  \right.
  \,.
\end{align}
Given a mass parameter $m_{200}$, a concentration parameter $c_{200}$
is evaluated in the simulations at the redshift $z$.  Since
$\sigma(M)$ is significantly altered by the bump compared to the
conventional case, we adopt the concentration-mass relation given by
Ref.~\cite{Sanchez-Conde:2013yxa}, where the fitting function is given
in terms of $\sigma(M)$.\footnote{See also
Refs.\,\cite{Correa:2015dva,Hu:2002we,Ishiyama:2011af} and Appendix B
of Ref.\,\cite{Hiroshima:2018kfv}.} On the other hand, $m_{200}$
relates to $r_{200}$ via $m_{200}=(4\pi/3)200\rho_c(z) r_{200}^3$,
where $\rho_c(z)$ is the critical density at the redshift $z$. Using
the relation $c_{200}=r_{200}/r_{\rm s}$, the scale radius $r_{\rm s}$
is determined. Then, $\rho_{\rm s}$ is obtained as
\begin{align}
  m &=\int d^3x\,\rho = 4\pi \rho_{\rm s}r_{\rm s}^3 f(c)\,, 
  \label{eq:mNFW}
\end{align}
where $f(c)=\ln(1+c)-c/(1+c)$ and $m=m_{200}$ and $c=c_{200}$ are
taken. We consider the $(r_{\rm s},\rho_{\rm s})$ as the values at the
accretion of a subhalo onto a host halo, which are denoted as
$(r_{{\rm s},a},\rho_{{\rm s},a})$. In order to discuss the tidal
stripping process of subhalo after the accretion, it is appropriate to
use the virial mass $m_{{\rm vir},a}$ instead of $m_{200}$. $m_{{\rm
    vir},a}$ is obtained by using Eq.\,\eqref{eq:mNFW} where
$m=m_{{\rm vir},a}$ and $c=c_{{\rm vir},a}=r_{{\rm vir},a}/r_{{\rm
    s},a}$, and by
\begin{align}
  m_{{\rm vir},a}
  &= \frac{4\pi}{3} \Delta_c(z_a) \rho_c(z_a) r_{{\rm vir},a}^3\,.
  \label{eq:ma1}
\end{align}
Here $\Delta_c(z)$ is given in Ref.\,\cite{Bryan:1997dn}.  To sum up,
we get the parameters $\rho_{{\rm s},a}$, $r_{{\rm s},a}$ and $m_{{\rm
    vir},a}$ at the accretion.

After accretion, the subhalo loses its mass due to the tidal
stripping. Given a value of the subhalo mass $m_0$ at the redshift
$z_0$ after tidal stripping, $(\rho_{{\rm s},a}, r_{{\rm s},a})$ are
translated into $(\rho_{{\rm s},0}, r_{{\rm s},0})$ at the redshift
$z_0$ using the relation among $V_{{\rm max},0}/V_{{\rm max},a}$,
$r_{{\rm max},0}/r_{{\rm max},a}$, and $m_0/m_{{\rm
    vir},a}$~\cite{Penarrubia_2010}. Here $V_{\rm max}$ and $r_{\rm
  max}$ are the maximum circular velocity and the radius which relate
to the NFW profile parameters as
\begin{align}
  V_{\rm max} = \sqrt{\frac{4\pi G \rho_{\rm s}}{4.625}}r_{\rm s}\,,
  ~~~r_{\rm max} = 2.163r_{\rm s}\,,
  \label{eq:Vmaxrmax}
\end{align}
where $G$ is the Newtonian constant. Using Eq.\,\eqref{eq:mNFW} with
$m=m_0$, $r_{\rm s}=r_{s,0}$, $\rho_{\rm s}=\rho_{{\rm s},0}$, and
$c_{t,0}=r_{t,0}/r_{{\rm s},0}$, we obtain the truncation radius
$r_{t,0}$ at $z=z_0$.  To summarize, given a subhalo mass $m_{{\rm
    vir},a}$ at the accretion and $m_0$ after tidal stripping, we
obtain subhalo properties, such as $\rho_{\rm s}$, $r_{\rm s}$, $r_t$,
and $V_{\rm max}$, which are important to discuss the observable
consequences.

The mass distribution function ${\cal F}$ of subhalos at the accretion
is given by the EPS formalism as a function of $m_a$, $z_a$ and the
host halo mass $M_0$ at $z=z_0$.  Then the number of subhalos with
mass $m_a$ that accrete at $z=z_a$ is given by
\begin{align}
  d^2N_{{\rm sh},a}=
{\cal F}(s_a,\delta_a|S_0,\delta_0) d\ln m_a d z_a\,,
\end{align}
where $s_a=\sigma^2(m_a)$, $\delta_a=\delta(z_a)$,
$S_0=\sigma^2(M_0)$, and $\delta_0=\delta(0)$. In the present
calculation, $M_0=M_{\rm host}(z=0)=1.3\times10^{12}M_\odot$ is taken.
Combining all the discussions above, the subhalo mass function after
the tidal stripping is obtained by
\begin{align}
  \frac{dN_{{\rm sh}}}{dm}=
  \int d^2N_{{\rm sh},a}
  \int dc_{{\rm vir},a} P^{c_{{\rm vir},a}}(m_a,z_{a}) \delta(m-m_0)\,,
\end{align}
where $P^{c_{{\rm vir},a}}(m_a,z_{a})$ is the distribution function
for $c_{{\rm vir},a}$ that is computed from the one for $c_{200}$-mass
relation discussed above Eq.\,\eqref{eq:mNFW}.

\section{Additional figures}
\label{app:figs}

We give additional figures with various values of $k_b$ for the
primordial curvature perturbation ${\cal P}_R(k)$
(Fig.\,\ref{fig:P_R}), the variance $\sigma(M)$ of the power spectrum
(Fig.\,\ref{fig:sigma}), and the average of the host halo mass
evolution $M_{\rm host}(z)$ (Fig.\,\ref{fig:hostM}).  The subhalo mass
function $dN_{\rm sh}/dm$ and the cumulative maximum circular velocity
function $N_{\rm sh}(>V_{\rm max})$ in the tidal model (a), (b), and
(c) are shown in Figs.\,\ref{fig:Nsh} and \ref{fig:NVmax},
respectively. Fig.\,\ref{fig:Nsh_map2} gives the color maps of the
cumulative number of subhalos the maximum circular velocity satisfying
$V_{\rm max}>4$~km/s, the number of dSphs whose mass is within
$10^5M_\odot$\,--\,$10^9M_\odot$ normalized by the one without the
bump, and the boost factor for the tidal model (a), (b), and (c).  The
mild dependence of the boost factor on $k_b$, which is mentioned in
the main text, is seen for all the tidal models.  This behavior can be
qualitatively understood from the results of the subhalo mass
function. In the plot of $dN_{\rm sh}/dm$, the bump is shifted to the
smaller scale as $k_b$ becomes large; meanwhile, the intensity of $m^2
dN_{\rm sh}/dm$ stays in the same order. This means that many subhalos
with lower masses are formed for bigger $k_b$, which leads to the
enhancement of the boost factor.

\begin{figure}[t]
  \begin{center}
    \includegraphics[scale=0.4]{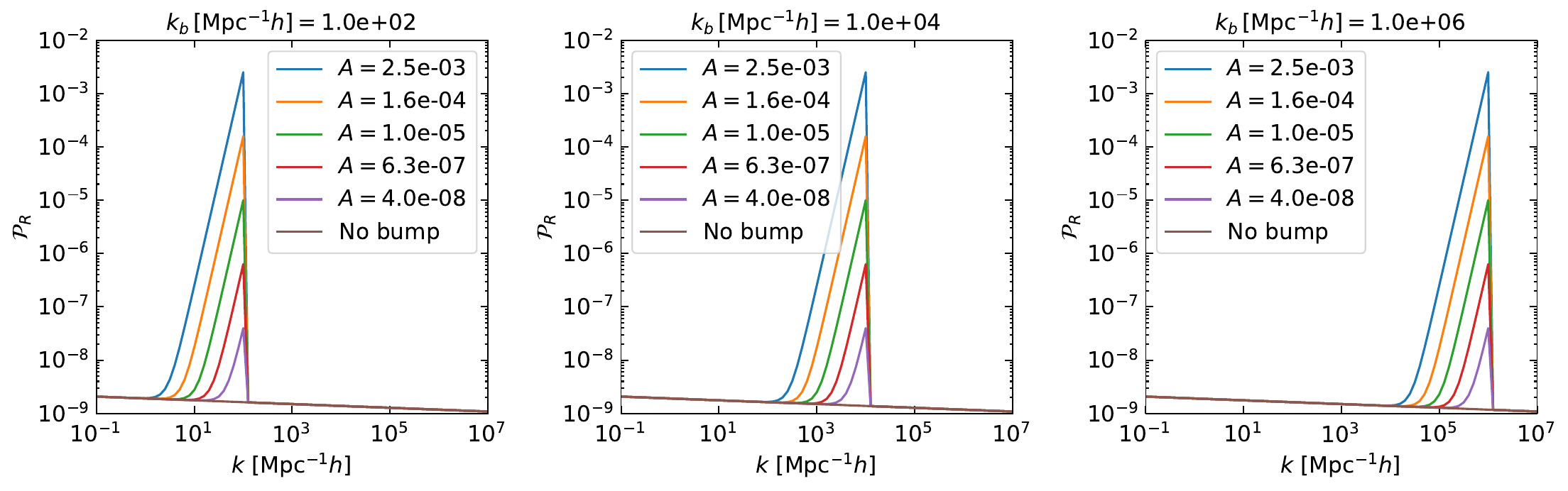}
  \end{center}
  \caption{Primordial curvature perturbation ${\cal P}_R$ as function
    of wavenumber. $k_b=1.0\times 10^{2}~{\rm Mpc}^{-1}h$ (left),
    $1.0\times 10^{4}~{\rm Mpc}^{-1}h$ (center), and $1.0\times
    10^{6}~{\rm Mpc}^{-1}h$ (right). Each line corresponds to
    $A=2.5\times 10^{-3}$, $1.6\times 10^{-4}$, $1.0\times 10^{-5}$,
    $6.3\times 10^{-7}$, and $4.0\times 10^{-8}$ from top to
    bottom. As a reference, ${\cal P}_R^{(0)}$ is shown as `No
    bump'. }
  \label{fig:P_R}
\end{figure}

\begin{figure}[t]
  \begin{center}
    \includegraphics[scale=0.4]{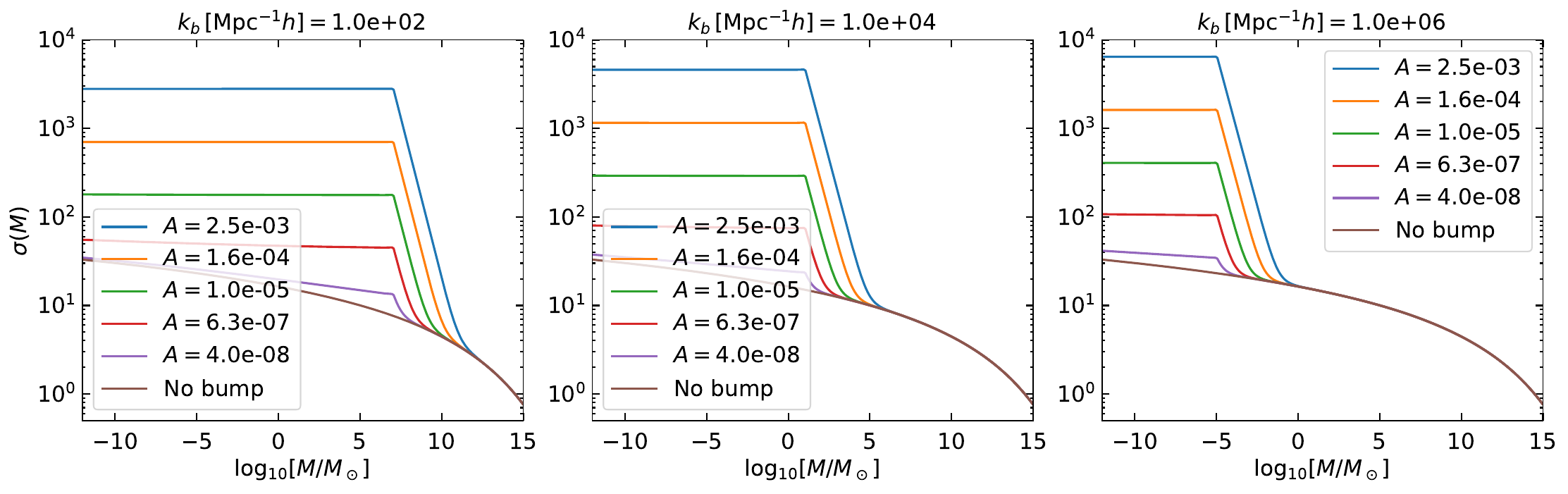}
  \end{center}
  \caption{Variance $\sigma$ of the power spectrum as functions of the
    mass scale $M$.  The parameters are the same as
    Fig.\,\ref{fig:P_R}. As a reference, $\sigma(M)$ without the bump
    is shown as `No bump', which is calculated by using
    CAMB~\cite{Lewis:1999bs}. }
  \label{fig:sigma}
\end{figure}

\begin{figure}[t]
  \begin{center}
    \includegraphics[scale=0.4]{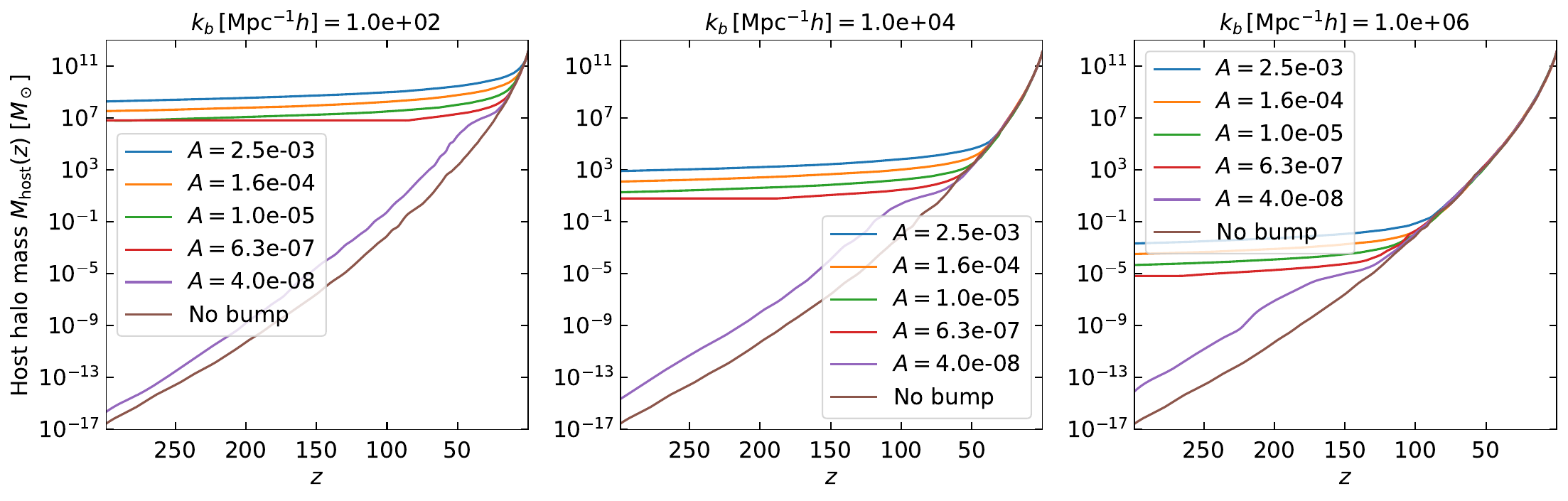}
  \end{center}
  \caption{Average of host halo mass evolution as functions of the
    redshift $z$. We take $M_{\rm host}(z=0)=1.3\times10^{12}M_\odot$
    and the other parameters are the same as
    Fig.\,\ref{fig:sigma}. The result without the bump is also shown
    as `No bump'.  }
  \label{fig:hostM}
\end{figure}

\begin{figure}[h]
  \begin{center}
    \includegraphics[scale=0.4]{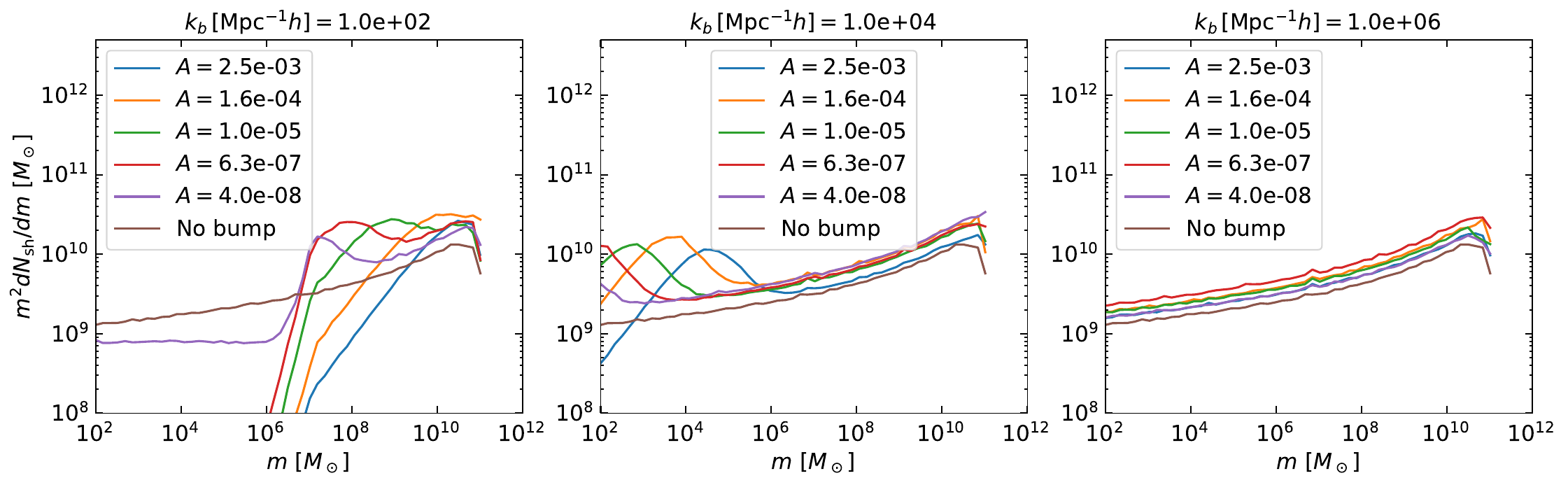}
    \includegraphics[scale=0.4]{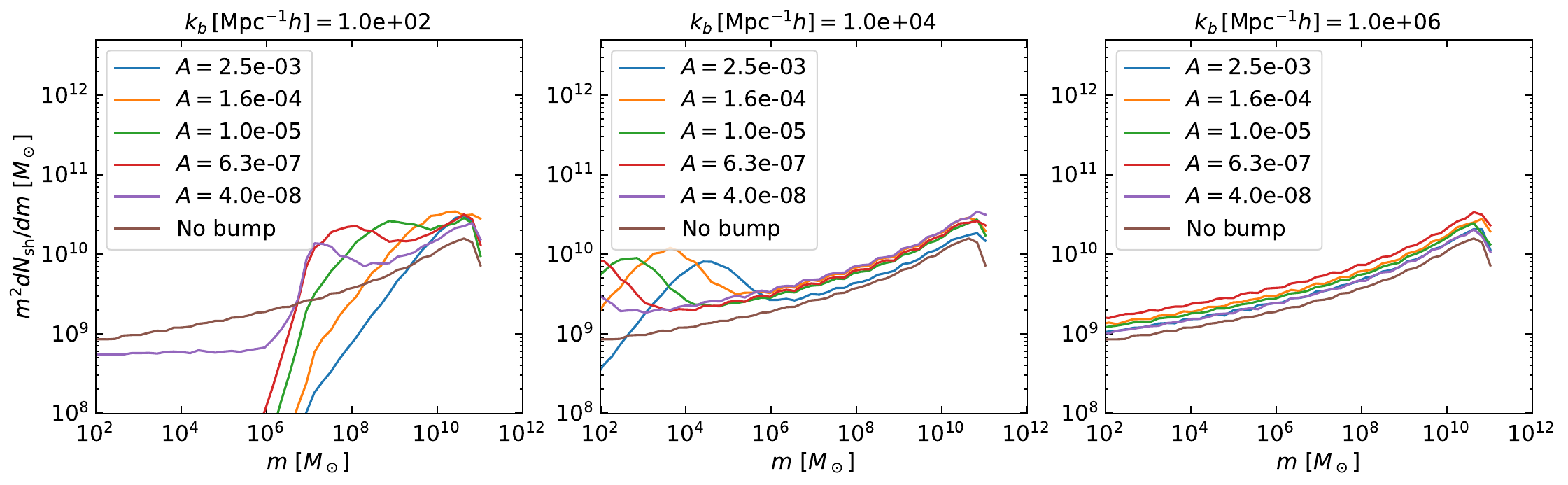}
    \includegraphics[scale=0.4]{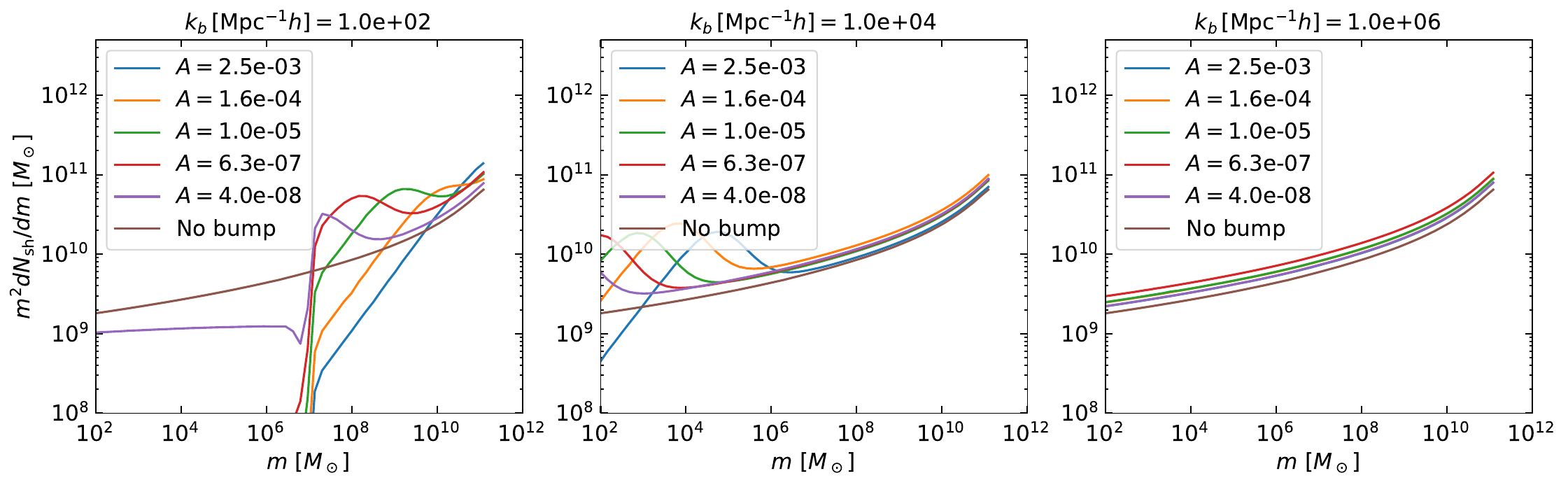}
  \end{center}
  \caption{Mass function of subhalo. Tidal model corresponds to (a)
    Jiang \& van den Bosch~\cite{Jiang:2014nsa}, (b) Hiroshima {\it et
      al.}~\cite{Hiroshima:2018kfv}, and (c) no tidal stripping, from top
    to bottom. The parameters are the same as Fig.\,\ref{fig:sigma}.
  }
  \label{fig:Nsh}
\end{figure}

\begin{figure}[t]
  \begin{center}
    \includegraphics[scale=0.4]{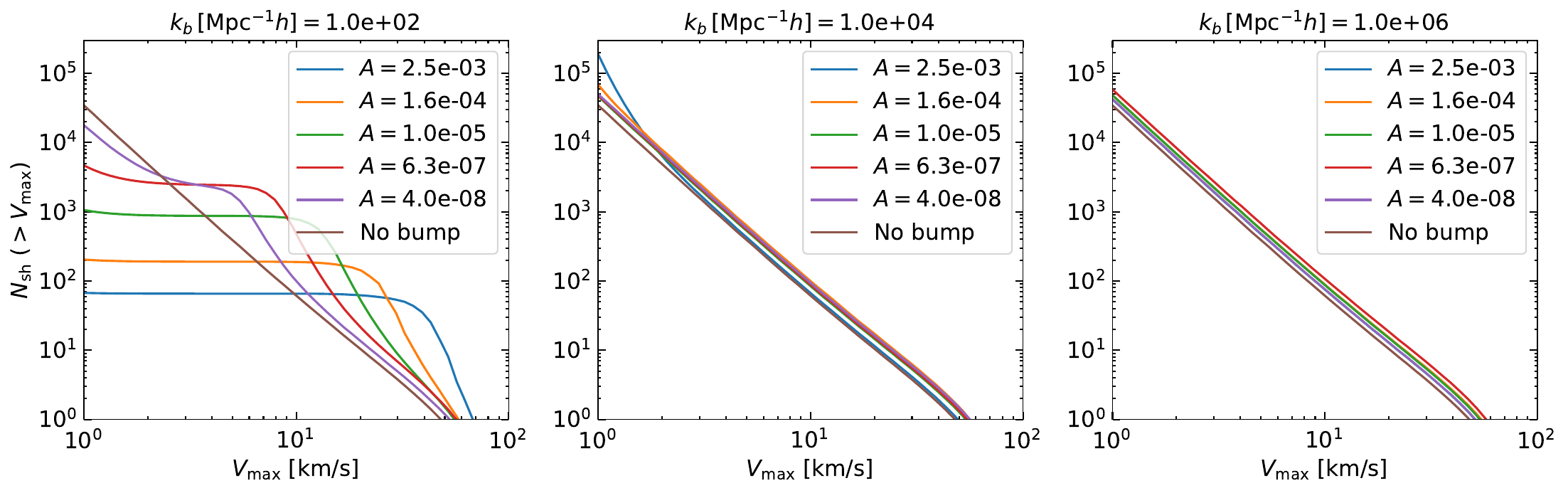}
    \includegraphics[scale=0.4]{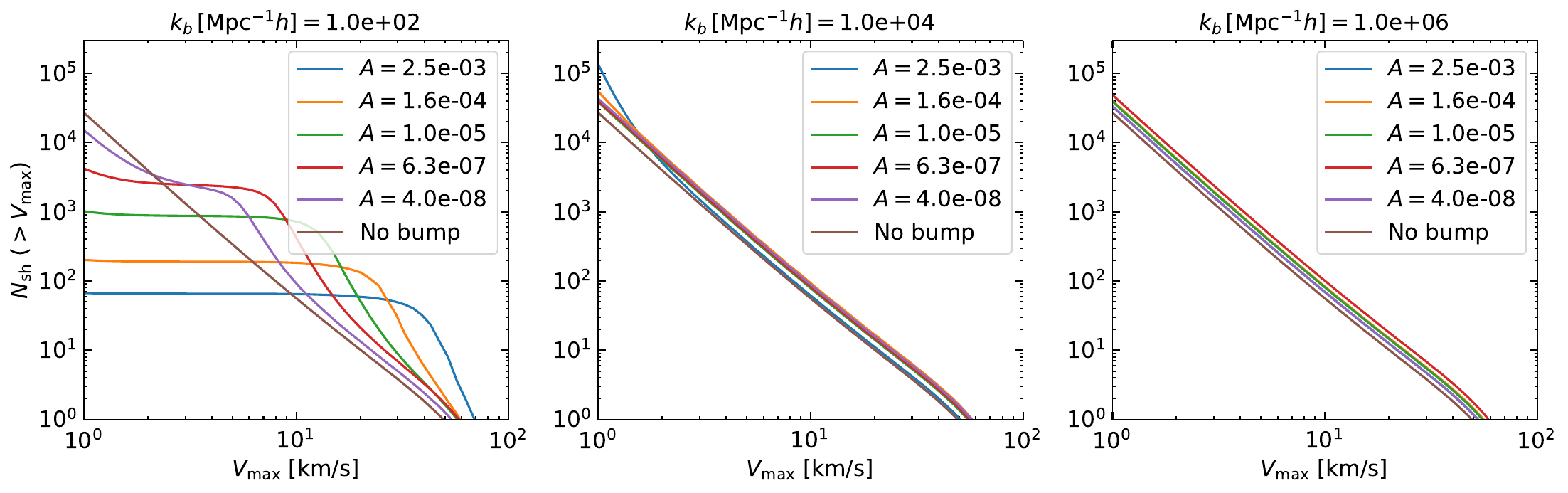}
    \includegraphics[scale=0.4]{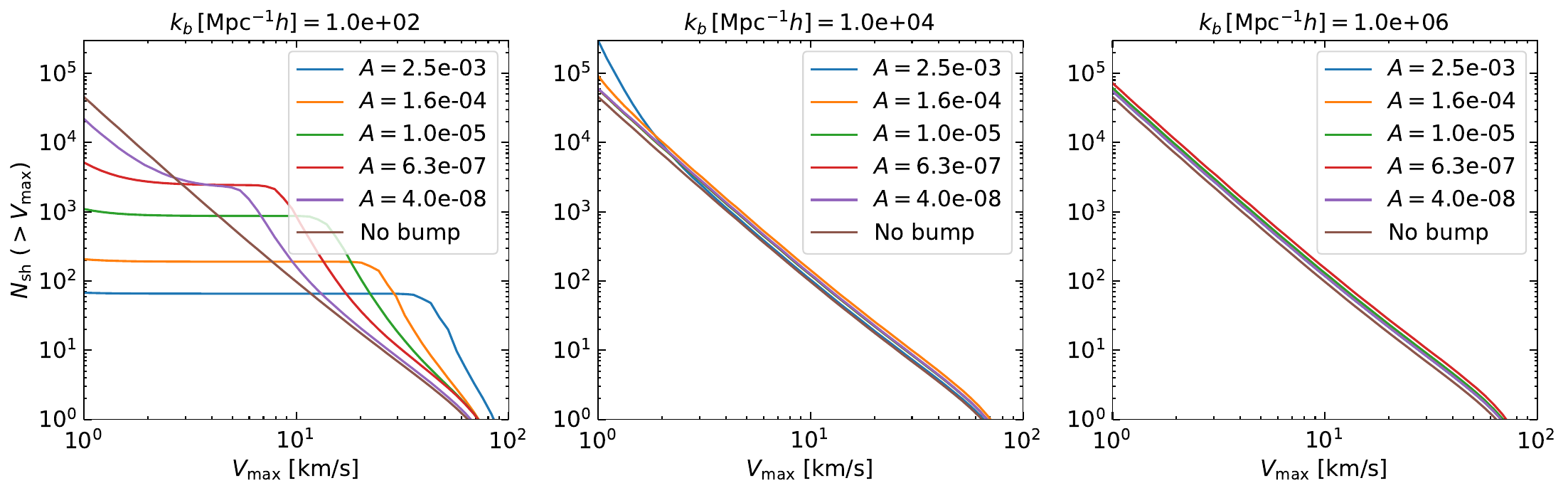}
  \end{center}
  \caption{Cumulative maximum circular velocity function. Tidal
    models, the parameters, and the ordering of panels are the same as
    Fig.\,\ref{fig:Nsh}.  }
  \label{fig:NVmax}
\end{figure}

\begin{figure}[t]
 \begin{center}
    \includegraphics[scale=0.17]{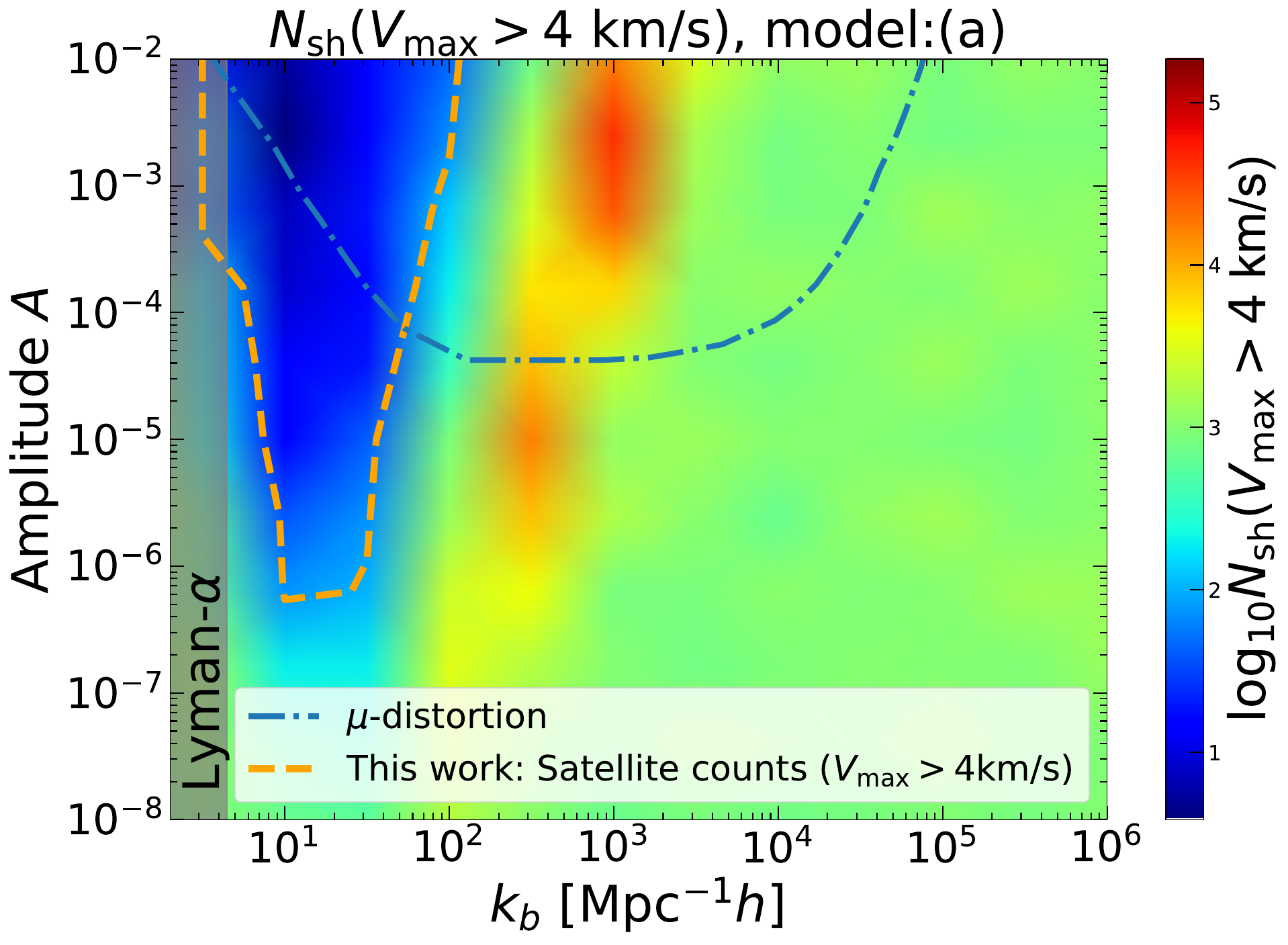}
    \includegraphics[scale=0.17]{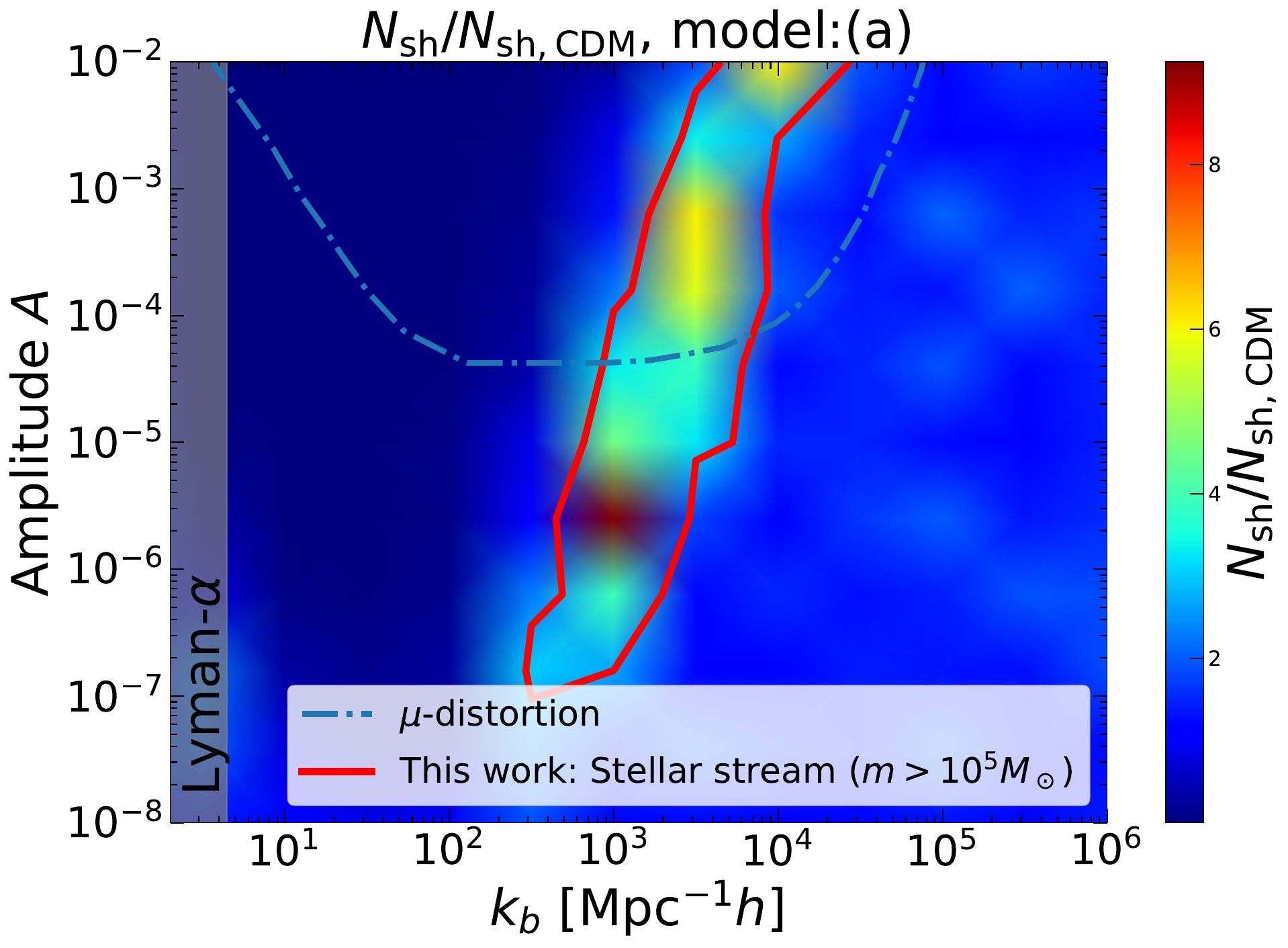}
    \includegraphics[scale=0.17]{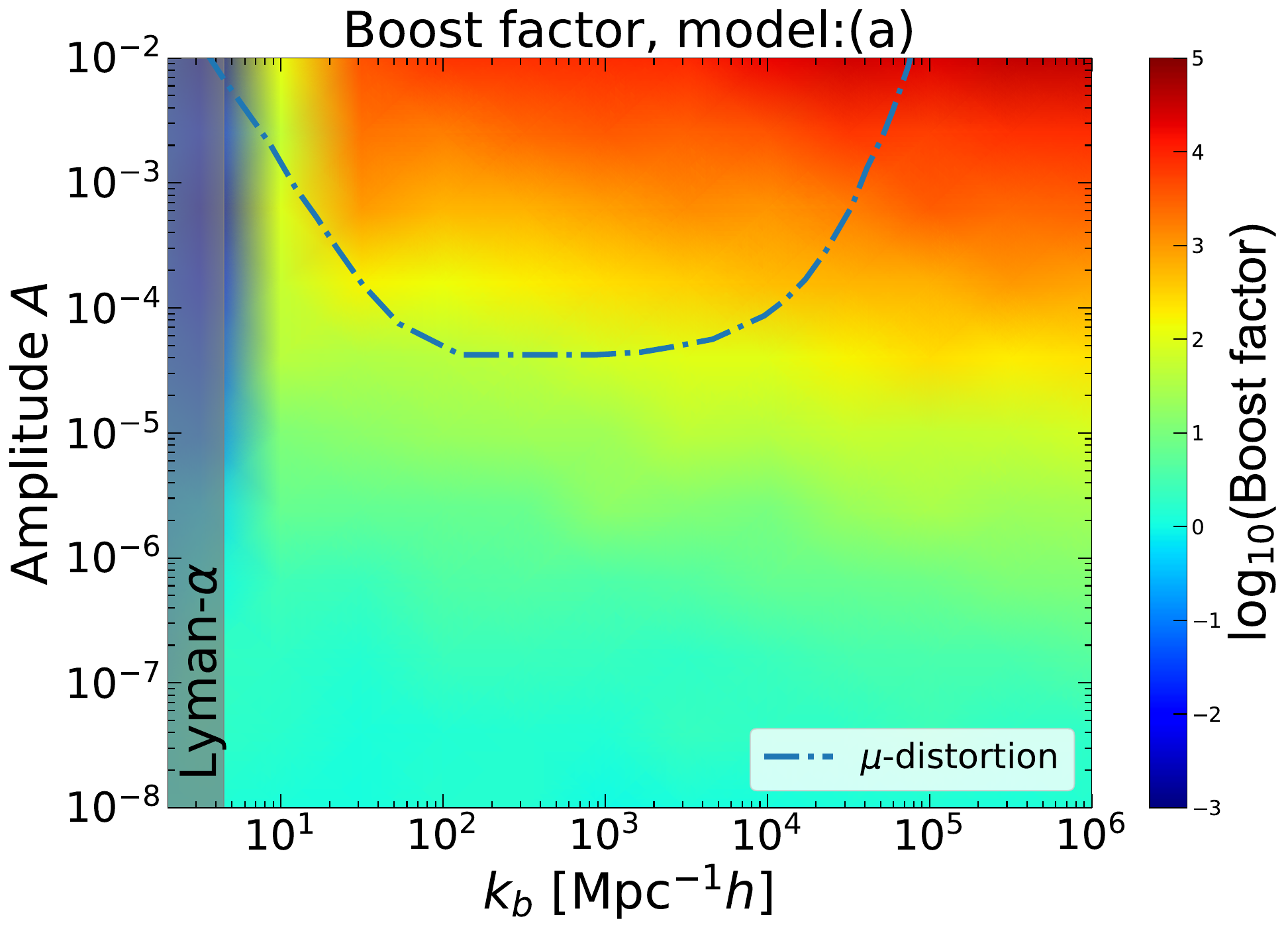}
    \includegraphics[scale=0.17]{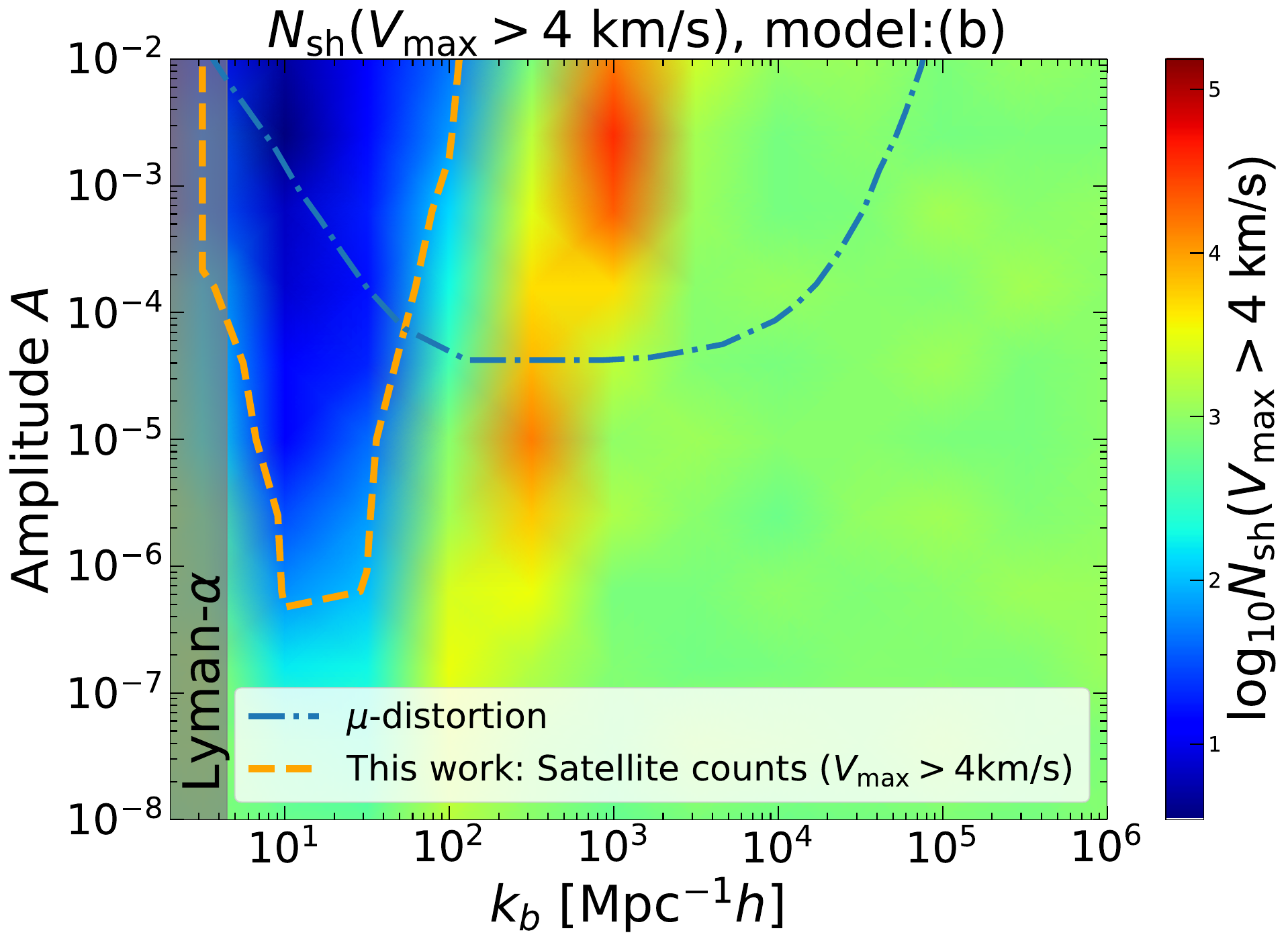}
    \includegraphics[scale=0.17]{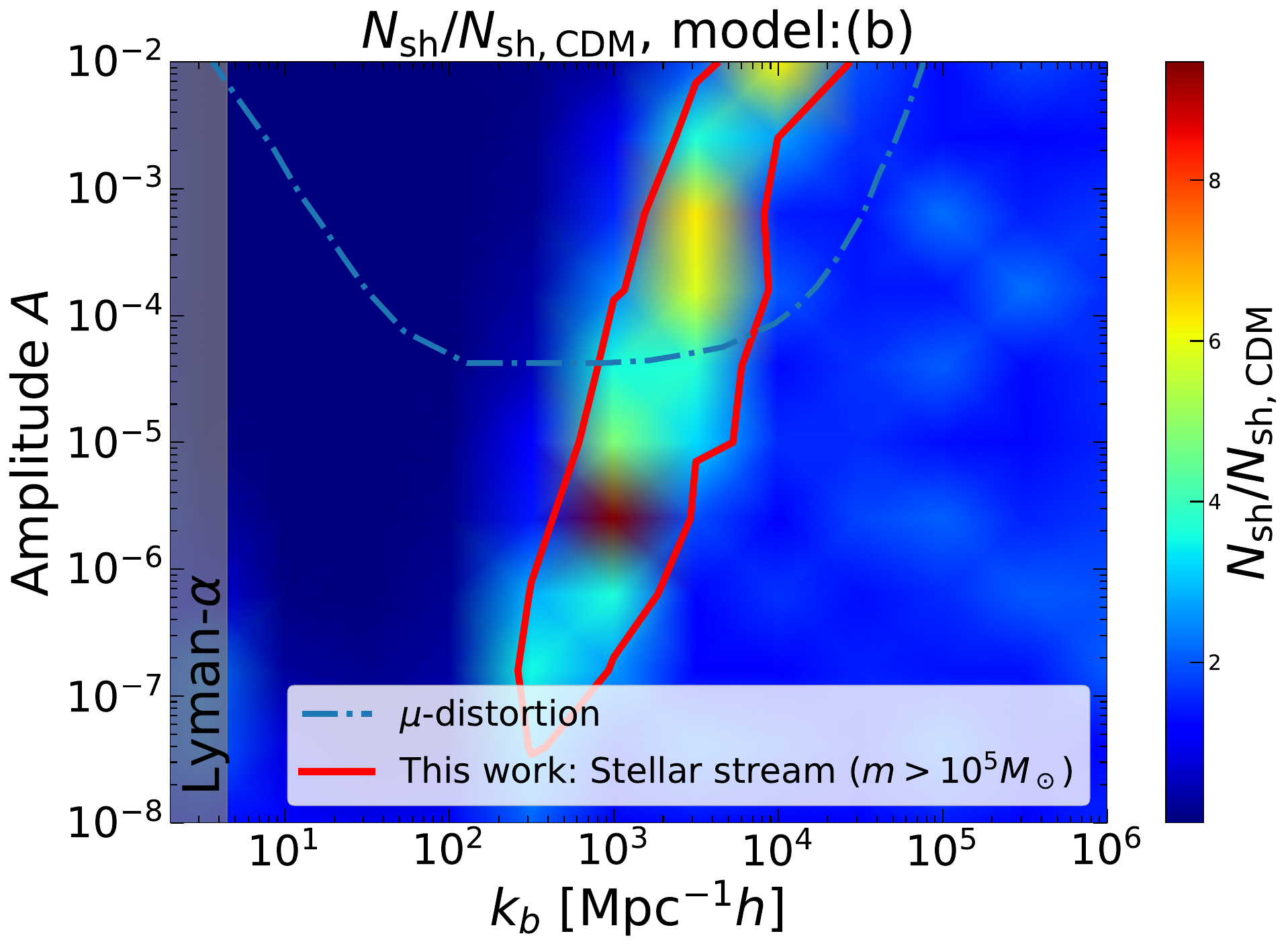}
    \includegraphics[scale=0.17]{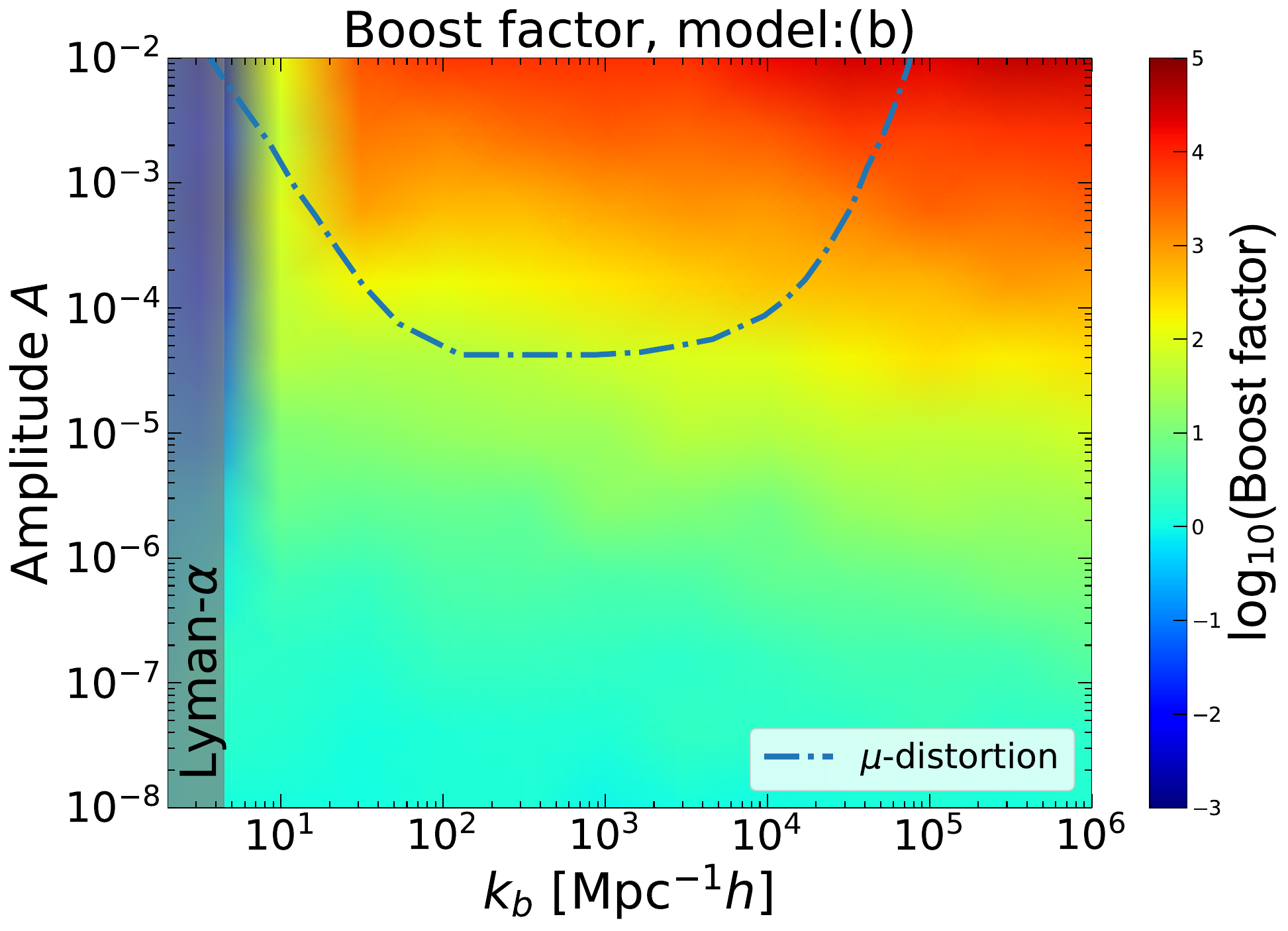}
    \includegraphics[scale=0.17]{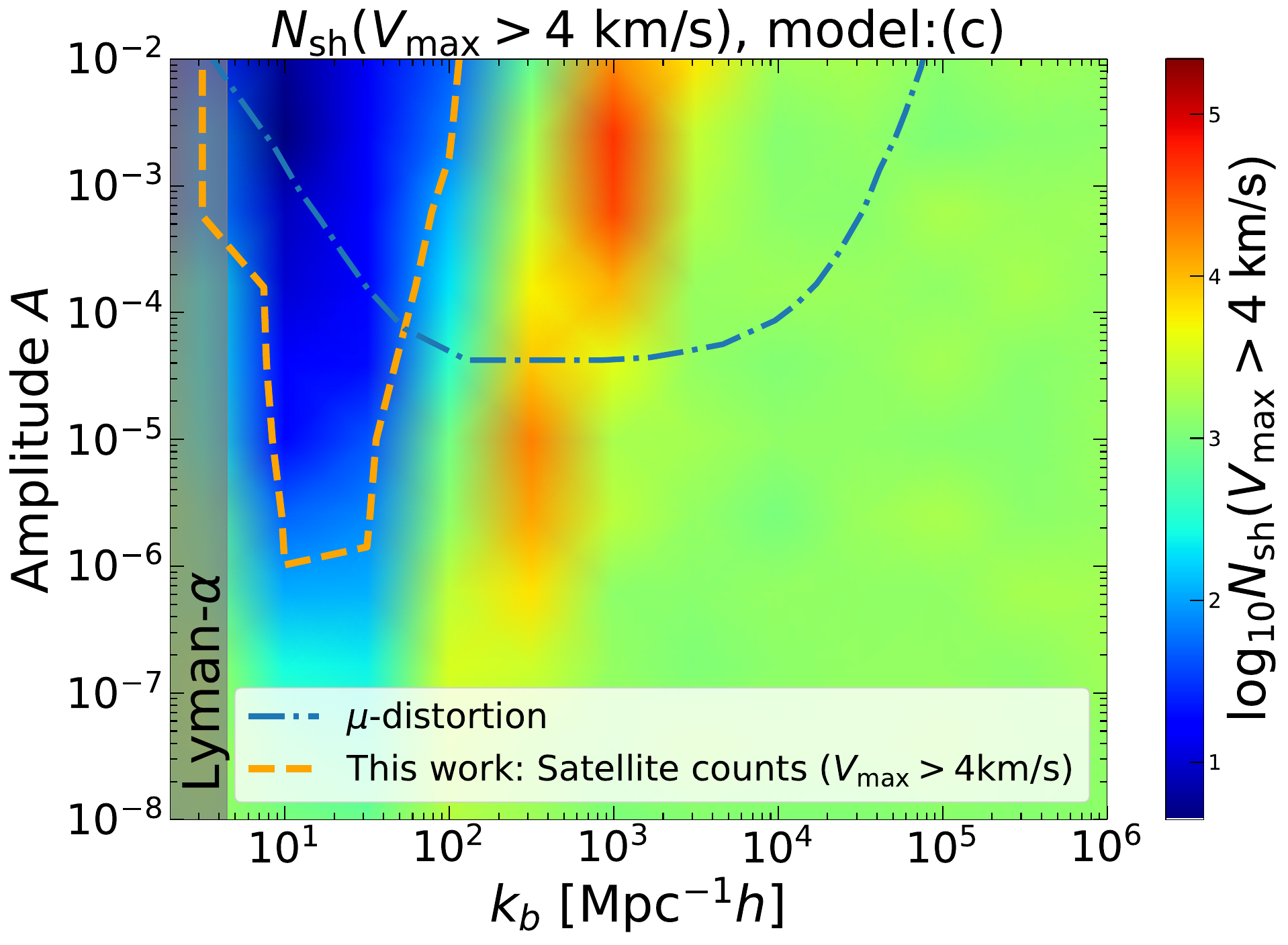}
    \includegraphics[scale=0.17]{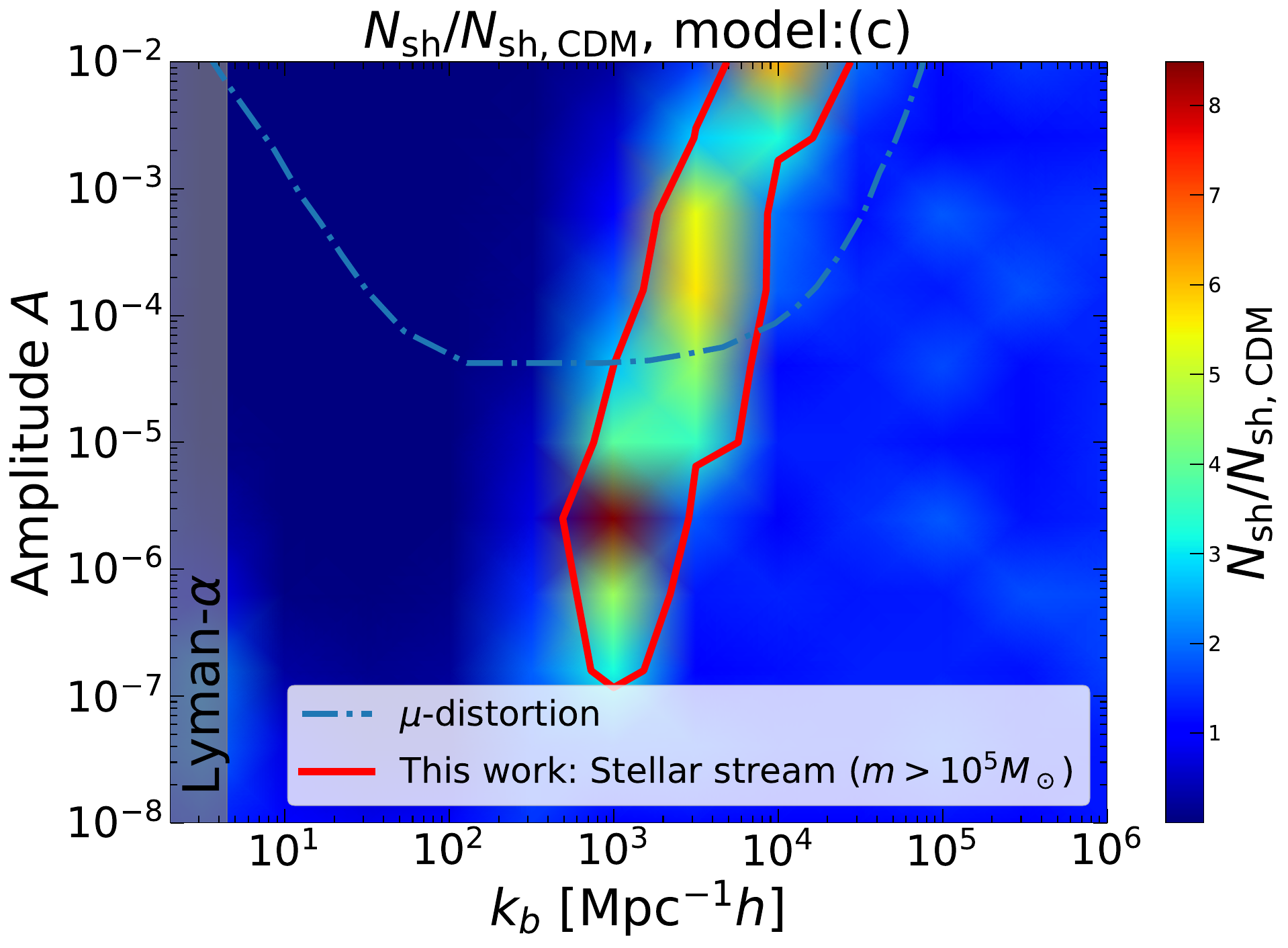}
    \includegraphics[scale=0.17]{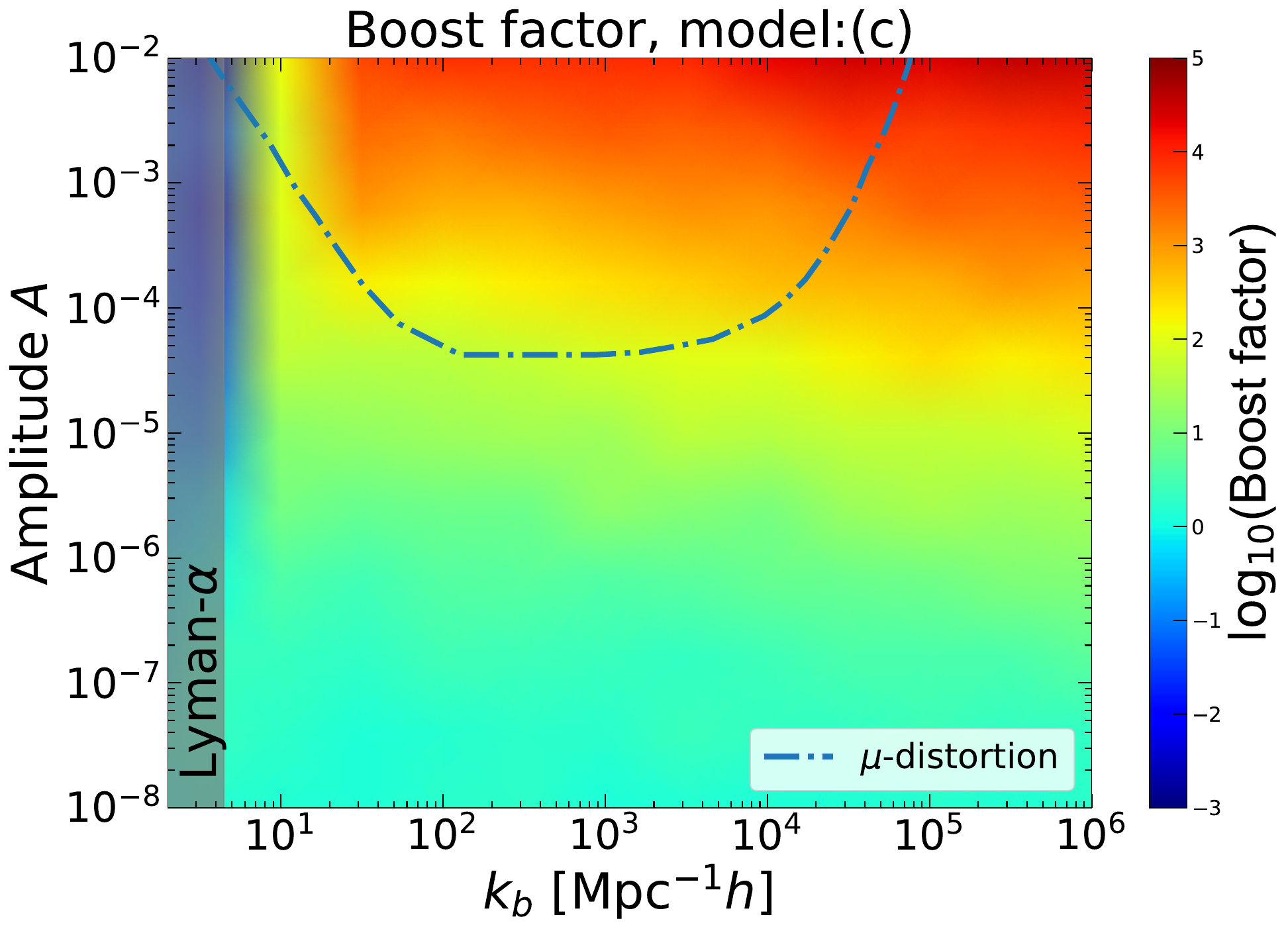}
  \end{center}
  \caption{Color map of the cumulative number of subhalos with the
    maximum circular velocity satisfying $V_{\rm max}>4$~km/s (left) ,
    $N_{\rm sh}/N_{\rm sh,CDM}$ (middle), and boost factor (right) on
    $(A,\,k_b)$ plane. For left and right panels, 95\% C.L. limits are
    also shown as in Fig.~\ref{fig:summary}.The tidal models are (a)
    Jiang \& van den Bosch~\cite{Jiang:2014nsa}, (b) Hiroshima {\it et
      al.}~\cite{Hiroshima:2018kfv}, and (c) no tidal stripping, from
    the top to the bottom. As a reference, the constraint due to
    $\mu$-distortion is shown as ``$\mu$-distortion'', which is given
    in Ref.\,\cite{Byrnes:2018txb}. Shaded region on the left is
    disfavored from the Lyman-$\alpha$
    observations~\cite{Bird:2010mp}. }
  \label{fig:Nsh_map2}
\end{figure}

\end{widetext}

\bibliographystyle{utphys}
\bibliography{refs}

\providecommand{\href}[2]{#2}\begingroup\raggedright\begin{thebibliography}{10}

\bibitem{Aghanim:2018eyx}
{\bfseries Planck} Collaboration, N.~Aghanim {\em et~al.}, ``{Planck 2018
  results. VI. Cosmological parameters},''
  \href{http://dx.doi.org/10.1051/0004-6361/201833910}{{\em Astron. Astrophys.}
  {\bfseries 641} (2020) A6}, \href{http://arxiv.org/abs/1807.06209}{{\ttfamily
  arXiv:1807.06209 [astro-ph.CO]}}.

\bibitem{Chluba:2012we}
J.~Chluba, A.~L. Erickcek, and I.~Ben-Dayan, ``{Probing the inflaton:
  Small-scale power spectrum constraints from measurements of the CMB energy
  spectrum},'' \href{http://dx.doi.org/10.1088/0004-637X/758/2/76}{{\em
  Astrophys. J.} {\bfseries 758} (2012) 76},
  \href{http://arxiv.org/abs/1203.2681}{{\ttfamily arXiv:1203.2681
  [astro-ph.CO]}}.

\bibitem{Chluba:2015bqa}
J.~Chluba, J.~Hamann, and S.~P. Patil, ``{Features and New Physical Scales in
  Primordial Observables: Theory and Observation},''
  \href{http://dx.doi.org/10.1142/S0218271815300232}{{\em Int. J. Mod. Phys. D}
  {\bfseries 24} no.~10, (2015) 1530023},
  \href{http://arxiv.org/abs/1505.01834}{{\ttfamily arXiv:1505.01834
  [astro-ph.CO]}}.

\bibitem{Josan:2009qn}
A.~S. Josan, A.~M. Green, and K.~A. Malik, ``{Generalised constraints on the
  curvature perturbation from primordial black holes},''
  \href{http://dx.doi.org/10.1103/PhysRevD.79.103520}{{\em Phys. Rev. D}
  {\bfseries 79} (2009) 103520},
  \href{http://arxiv.org/abs/0903.3184}{{\ttfamily arXiv:0903.3184
  [astro-ph.CO]}}.

\bibitem{Carr:2009jm}
B.~J. Carr, K.~Kohri, Y.~Sendouda, and J.~Yokoyama, ``{New cosmological
  constraints on primordial black holes},''
  \href{http://dx.doi.org/10.1103/PhysRevD.81.104019}{{\em Phys. Rev. D}
  {\bfseries 81} (2010) 104019},
  \href{http://arxiv.org/abs/0912.5297}{{\ttfamily arXiv:0912.5297
  [astro-ph.CO]}}.

\bibitem{Byrnes:2018txb}
C.~T. Byrnes, P.~S. Cole, and S.~P. Patil, ``{Steepest growth of the power
  spectrum and primordial black holes},''
  \href{http://dx.doi.org/10.1088/1475-7516/2019/06/028}{{\em JCAP} {\bfseries
  06} (2019) 028}, \href{http://arxiv.org/abs/1811.11158}{{\ttfamily
  arXiv:1811.11158 [astro-ph.CO]}}.

\bibitem{Dalianis:2018ymb}
I.~Dalianis, ``{Constraints on the curvature power spectrum from primordial
  black hole evaporation},''
  \href{http://dx.doi.org/10.1088/1475-7516/2019/08/032}{{\em JCAP} {\bfseries
  08} (2019) 032}, \href{http://arxiv.org/abs/1812.09807}{{\ttfamily
  arXiv:1812.09807 [astro-ph.CO]}}.

\bibitem{Sato-Polito:2019hws}
G.~Sato-Polito, E.~D. Kovetz, and M.~Kamionkowski, ``{Constraints on the
  primordial curvature power spectrum from primordial black holes},''
  \href{http://dx.doi.org/10.1103/PhysRevD.100.063521}{{\em Phys. Rev. D}
  {\bfseries 100} no.~6, (2019) 063521},
  \href{http://arxiv.org/abs/1904.10971}{{\ttfamily arXiv:1904.10971
  [astro-ph.CO]}}.

\bibitem{Gow:2020bzo}
A.~D. Gow, C.~T. Byrnes, P.~S. Cole, and S.~Young, ``{The power spectrum on
  small scales: Robust constraints and comparing PBH methodologies},''
  \href{http://dx.doi.org/10.1088/1475-7516/2021/02/002}{{\em JCAP} {\bfseries
  02} (2021) 002}, \href{http://arxiv.org/abs/2008.03289}{{\ttfamily
  arXiv:2008.03289 [astro-ph.CO]}}.

\bibitem{Delos:2018ueo}
M.~S. Delos, A.~L. Erickcek, A.~P. Bailey, and M.~A. Alvarez, ``{Density
  profiles of ultracompact minihalos: Implications for constraining the
  primordial power spectrum},''
  \href{http://dx.doi.org/10.1103/PhysRevD.98.063527}{{\em Phys. Rev. D}
  {\bfseries 98} no.~6, (2018) 063527},
  \href{http://arxiv.org/abs/1806.07389}{{\ttfamily arXiv:1806.07389
  [astro-ph.CO]}}.

\bibitem{Nakama:2017qac}
T.~Nakama, T.~Suyama, K.~Kohri, and N.~Hiroshima, ``{Constraints on small-scale
  primordial power by annihilation signals from extragalactic dark matter
  minihalos},'' \href{http://dx.doi.org/10.1103/PhysRevD.97.023539}{{\em Phys.
  Rev. D} {\bfseries 97} no.~2, (2018) 023539},
  \href{http://arxiv.org/abs/1712.08820}{{\ttfamily arXiv:1712.08820
  [astro-ph.CO]}}.

\bibitem{Abe:2021mcv}
K.~T. Abe, T.~Minoda, and H.~Tashiro, ``{Constraint on the early-formed dark
  matter halos using the free-free emission in the Planck foreground
  analysis},'' \href{http://dx.doi.org/10.1103/PhysRevD.105.063531}{{\em Phys.
  Rev. D} {\bfseries 105} no.~6, (2022) 063531},
  \href{http://arxiv.org/abs/2108.00621}{{\ttfamily arXiv:2108.00621
  [astro-ph.CO]}}.

\bibitem{Yoshiura:2020soa}
S.~Yoshiura, M.~Oguri, K.~Takahashi, and T.~Takahashi, ``{Constraints on
  primordial power spectrum from galaxy luminosity functions},''
  \href{http://dx.doi.org/10.1103/PhysRevD.102.083515}{{\em Phys. Rev. D}
  {\bfseries 102} no.~8, (2020) 083515},
  \href{http://arxiv.org/abs/2007.14695}{{\ttfamily arXiv:2007.14695
  [astro-ph.CO]}}.

\bibitem{Sabti:2021xvh}
N.~Sabti, J.~B. Mu\~noz, and D.~Blas, ``{Galaxy luminosity function pipeline
  for cosmology and astrophysics},''
  \href{http://dx.doi.org/10.1103/PhysRevD.105.043518}{{\em Phys. Rev. D}
  {\bfseries 105} no.~4, (2022) 043518},
  \href{http://arxiv.org/abs/2110.13168}{{\ttfamily arXiv:2110.13168
  [astro-ph.CO]}}.

\bibitem{Gilman:2021gkj}
D.~Gilman, A.~Benson, J.~Bovy, S.~Birrer, T.~Treu, and A.~Nierenberg, ``{The
  primordial matter power spectrum on sub-galactic scales},''
  \href{http://dx.doi.org/10.1093/mnras/stac670}{{\em Mon. Not. Roy. Astron.
  Soc.} {\bfseries 512} no.~3, (2022) 3163--3188},
  \href{http://arxiv.org/abs/2112.03293}{{\ttfamily arXiv:2112.03293
  [astro-ph.CO]}}.

\bibitem{Fermi-LAT:2016uux}
{\bfseries Fermi-LAT, DES} Collaboration, A.~Albert {\em et~al.}, ``{Searching
  for Dark Matter Annihilation in Recently Discovered Milky Way Satellites with
  Fermi-LAT},'' \href{http://dx.doi.org/10.3847/1538-4357/834/2/110}{{\em
  Astrophys. J.} {\bfseries 834} no.~2, (2017) 110},
  \href{http://arxiv.org/abs/1611.03184}{{\ttfamily arXiv:1611.03184
  [astro-ph.HE]}}.

\bibitem{Hoof:2018hyn}
S.~Hoof, A.~Geringer-Sameth, and R.~Trotta, ``{A Global Analysis of Dark Matter
  Signals from 27 Dwarf Spheroidal Galaxies using 11 Years of Fermi-LAT
  Observations},'' \href{http://dx.doi.org/10.1088/1475-7516/2020/02/012}{{\em
  JCAP} {\bfseries 02} (2020) 012},
  \href{http://arxiv.org/abs/1812.06986}{{\ttfamily arXiv:1812.06986
  [astro-ph.CO]}}.

\bibitem{Ando:2020yyk}
S.~Ando, A.~Geringer-Sameth, N.~Hiroshima, S.~Hoof, R.~Trotta, and M.~G.
  Walker, ``{Structure formation models weaken limits on WIMP dark matter from
  dwarf spheroidal galaxies},''
  \href{http://dx.doi.org/10.1103/PhysRevD.102.061302}{{\em Phys. Rev. D}
  {\bfseries 102} no.~6, (2020) 061302},
  \href{http://arxiv.org/abs/2002.11956}{{\ttfamily arXiv:2002.11956
  [astro-ph.CO]}}.

\bibitem{Bird:2010mp}
S.~Bird, H.~V. Peiris, M.~Viel, and L.~Verde, ``{Minimally Parametric Power
  Spectrum Reconstruction from the Lyman-alpha Forest},''
  \href{http://dx.doi.org/10.1111/j.1365-2966.2011.18245.x}{{\em Mon. Not. Roy.
  Astron. Soc.} {\bfseries 413} (2011) 1717--1728},
  \href{http://arxiv.org/abs/1010.1519}{{\ttfamily arXiv:1010.1519
  [astro-ph.CO]}}.

\bibitem{Hiroshima:2018kfv}
N.~Hiroshima, S.~Ando, and T.~Ishiyama, ``{Modeling evolution of dark matter
  substructure and annihilation boost},''
  \href{http://dx.doi.org/10.1103/PhysRevD.97.123002}{{\em Phys. Rev. D}
  {\bfseries 97} no.~12, (2018) 123002},
  \href{http://arxiv.org/abs/1803.07691}{{\ttfamily arXiv:1803.07691
  [astro-ph.CO]}}.

\bibitem{Ando:2019xlm}
S.~Ando, T.~Ishiyama, and N.~Hiroshima, ``{Halo Substructure Boosts to the
  Signatures of Dark Matter Annihilation},''
  \href{http://dx.doi.org/10.3390/galaxies7030068}{{\em Galaxies} {\bfseries 7}
  no.~3, (2019) 68}, \href{http://arxiv.org/abs/1903.11427}{{\ttfamily
  arXiv:1903.11427 [astro-ph.CO]}}.

\bibitem{Simon:2019nxf}
J.~D. Simon, ``{The Faintest Dwarf Galaxies},''
  \href{http://dx.doi.org/10.1146/annurev-astro-091918-104453}{{\em Ann. Rev.
  Astron. Astrophys.} {\bfseries 57} no.~1, (2019) 375--415},
  \href{http://arxiv.org/abs/1901.05465}{{\ttfamily arXiv:1901.05465
  [astro-ph.GA]}}.

\bibitem{Collins:2017}
M.~L.~M. Collins, E.~J. Tollerud, D.~J. Sand, A.~Bonaca, B.~Willman, and
  J.~Strader, ``{Dynamical evidence for a strong tidal interaction between the
  Milky Way and its satellite, Leo V},''
  \href{http://dx.doi.org/10.1093/mnras/stx067}{{\em Mon. Not. Roy. Astron.
  Soc.} {\bfseries 467} no.~1, (2017) 573--585},
  \href{http://arxiv.org/abs/1608.05710}{{\ttfamily arXiv:1608.05710
  [astro-ph.GA]}}.

\bibitem{Banik:2019cza}
N.~Banik, J.~Bovy, G.~Bertone, D.~Erkal, and T.~J.~L. de~Boer, ``{Evidence of a
  population of dark subhaloes from $Gaia$ and Pan-STARRS observations of the
  GD-1 stream},'' \href{http://dx.doi.org/10.1093/mnras/stab210}{{\em Mon. Not.
  Roy. Astron. Soc.} {\bfseries 502} no.~2, (2021) 2364--2380},
  \href{http://arxiv.org/abs/1911.02662}{{\ttfamily arXiv:1911.02662
  [astro-ph.GA]}}.

\bibitem{Grillmair:2006bd}
C.~J. Grillmair and O.~Dionatos, ``{Detection of a 63 Degree Cold Stellar
  Stream in the Sloan Digital Sky Survey},''
  \href{http://dx.doi.org/10.1086/505111}{{\em Astrophys. J. Lett.} {\bfseries
  643} (2006) L17--L20},
  \href{http://arxiv.org/abs/astro-ph/0604332}{{\ttfamily
  arXiv:astro-ph/0604332}}.

\bibitem{Dalal:2002su}
N.~Dalal and C.~S. Kochanek, ``{Strong lensing constraints on small scale
  linear power},'' \href{http://arxiv.org/abs/astro-ph/0202290}{{\ttfamily
  arXiv:astro-ph/0202290}}.

\bibitem{Ozsoy:2019lyy}
O.~Ozsoy and G.~Tasinato, ``{On the slope of the curvature power spectrum in
  non-attractor inflation},''
  \href{http://dx.doi.org/10.1088/1475-7516/2020/04/048}{{\em JCAP} {\bfseries
  04} (2020) 048}, \href{http://arxiv.org/abs/1912.01061}{{\ttfamily
  arXiv:1912.01061 [astro-ph.CO]}}.

\bibitem{Schneider:2013ria}
A.~Schneider, R.~E. Smith, and D.~Reed, ``{Halo Mass Function and the Free
  Streaming Scale},'' \href{http://dx.doi.org/10.1093/mnras/stt829}{{\em Mon.
  Not. Roy. Astron. Soc.} {\bfseries 433} (2013) 1573},
  \href{http://arxiv.org/abs/1303.0839}{{\ttfamily arXiv:1303.0839
  [astro-ph.CO]}}.

\bibitem{Kadota:2020ahr}
K.~Kadota and J.~Silk, ``{Boosting small-scale structure via primordial black
  holes and implications for sub-GeV dark matter annihilation},''
  \href{http://dx.doi.org/10.1103/PhysRevD.103.043530}{{\em Phys. Rev. D}
  {\bfseries 103} no.~4, (2021) 043530},
  \href{http://arxiv.org/abs/2012.03698}{{\ttfamily arXiv:2012.03698
  [astro-ph.CO]}}.

\bibitem{Lacey:1993iv}
C.~G. Lacey and S.~Cole, ``{Merger rates in hierarchical models of galaxy
  formation},'' {\em Mon. Not. Roy. Astron. Soc.} {\bfseries 262} (1993)
  627--649.

\bibitem{Yang_2011}
X.~Yang, H.~J. Mo, Y.~Zhang, and F.~C. van~den Bosch, ``{AN} {ANALYTICAL}
  {MODEL} {FOR} {THE} {ACCRETION} {OF} {DARK} {MATTER} {SUBHALOS},''
  \href{http://dx.doi.org/10.1088/0004-637x/741/1/13}{{\em The Astrophysical
  Journal} {\bfseries 741} no.~1, (Oct, 2011) 13}.
  \url{https://doi.org/10.1088%2F0004-637x%2F741%2F1%2F13}.

\bibitem{Correa:2014xma}
C.~A. Correa, J.~S.~B. Wyithe, J.~Schaye, and A.~R. Duffy, ``{The accretion
  history of dark matter haloes \textendash{} I. The physical origin of the
  universal function},'' \href{http://dx.doi.org/10.1093/mnras/stv689}{{\em
  Mon. Not. Roy. Astron. Soc.} {\bfseries 450} no.~2, (2015) 1514--1520},
  \href{http://arxiv.org/abs/1409.5228}{{\ttfamily arXiv:1409.5228
  [astro-ph.GA]}}.

\bibitem{Sanchez-Conde:2013yxa}
M.~A. S\'anchez-Conde and F.~Prada, ``{The flattening of the
  concentration\textendash{}mass relation towards low halo masses and its
  implications for the annihilation signal boost},''
  \href{http://dx.doi.org/10.1093/mnras/stu1014}{{\em Mon. Not. Roy. Astron.
  Soc.} {\bfseries 442} no.~3, (2014) 2271--2277},
  \href{http://arxiv.org/abs/1312.1729}{{\ttfamily arXiv:1312.1729
  [astro-ph.CO]}}.

\bibitem{Navarro:1995iw}
J.~F. Navarro, C.~S. Frenk, and S.~D.~M. White, ``{The Structure of cold dark
  matter halos},'' \href{http://dx.doi.org/10.1086/177173}{{\em Astrophys. J.}
  {\bfseries 462} (1996) 563--575},
  \href{http://arxiv.org/abs/astro-ph/9508025}{{\ttfamily
  arXiv:astro-ph/9508025}}.

\bibitem{Jiang:2014nsa}
F.~Jiang and F.~C. van~den Bosch, ``{Statistics of dark matter substructure
  \textendash{} I. Model and universal fitting functions},''
  \href{http://dx.doi.org/10.1093/mnras/stw439}{{\em Mon. Not. Roy. Astron.
  Soc.} {\bfseries 458} no.~3, (2016) 2848--2869},
  \href{http://arxiv.org/abs/1403.6827}{{\ttfamily arXiv:1403.6827
  [astro-ph.CO]}}.

\bibitem{Delos:2019lik}
M.~S. Delos, ``{Tidal evolution of dark matter annihilation rates in
  subhalos},'' \href{http://dx.doi.org/10.1103/PhysRevD.100.063505}{{\em Phys.
  Rev. D} {\bfseries 100} no.~6, (2019) 063505},
  \href{http://arxiv.org/abs/1906.10690}{{\ttfamily arXiv:1906.10690
  [astro-ph.CO]}}.

\bibitem{Graus_2019}
A.~S. Graus, J.~S. Bullock, T.~Kelley, M.~Boylan-Kolchin, S.~Garrison-Kimmel,
  and Y.~Qi, ``How low does it go? too few galactic satellites with standard
  reionization quenching,'' \href{http://dx.doi.org/10.1093/mnras/stz1992}{{\em
  Monthly Notices of the Royal Astronomical Society} {\bfseries 488} no.~4,
  (Jul, 2019) 4585--4595}. \url{https://doi.org/10.1093%2Fmnras%2Fstz1992}.

\bibitem{Dekker:2021scf}
A.~Dekker, S.~Ando, C.~A. Correa, and K.~C.~Y. Ng, ``{Warm Dark Matter
  Constraints Using Milky-Way Satellite Observations and Subhalo Evolution
  Modeling},'' \href{http://arxiv.org/abs/2111.13137}{{\ttfamily
  arXiv:2111.13137 [astro-ph.CO]}}.

\bibitem{DES:2019vzn}
{\bfseries DES} Collaboration, A.~Drlica-Wagner {\em et~al.}, ``{Milky Way
  Satellite Census. I. The Observational Selection Function for Milky Way
  Satellites in DES Y3 and Pan-STARRS DR1},''
  \href{http://dx.doi.org/10.3847/1538-4357/ab7eb9}{{\em Astrophys. J.}
  {\bfseries 893} (2020) 47}, \href{http://arxiv.org/abs/1912.03302}{{\ttfamily
  arXiv:1912.03302 [astro-ph.GA]}}.

\bibitem{Gaia:2016}
{\bfseries Gaia} Collaboration, A.~G.~A. Brown {\em et~al.}, ``{Gaia Data
  Release 1. Summary of the astrometric, photometric, and survey properties},''
  \href{http://dx.doi.org/10.1051/0004-6361/201629512}{{\em Astronomy {\&}
  Astrophysics} {\bfseries 595} (2016) A2}.

\bibitem{Gaia:2018}
{\bfseries Gaia} Collaboration, L.~Lindegren {\em et~al.}, ``{Gaia Data Release
  2: The astrometric solution},''
  \href{http://dx.doi.org/10.48550/arXiv.1804.09366}{{\em Astronomy {\&}
  Astrophysics} {\bfseries 616} (2018) A2}.

\bibitem{Pan-STARRA1:2016}
{\bfseries Pan-STARRA1} Collaboration, K.~C. Chambers {\em et~al.}, ``{The
  Pan-STARRS1 Surveys},'' \href{http://arxiv.org/abs/1612.05560}{{\ttfamily
  arXiv:1612.05560 [astro-ph.IM]}}.

\bibitem{Diemand:2005vz}
J.~Diemand, B.~Moore, and J.~Stadel, ``{Earth-mass dark-matter haloes as the
  first structures in the early Universe},''
  \href{http://dx.doi.org/10.1038/nature03270}{{\em Nature} {\bfseries 433}
  (2005) 389--391}, \href{http://arxiv.org/abs/astro-ph/0501589}{{\ttfamily
  arXiv:astro-ph/0501589}}.

\bibitem{Wang:2022spb}
Q.~Wang, L.~Gao, and C.~Meng, ``{The Ultramarine Simulation: properties of dark
  matter haloes before redshift 5.5},''
  \href{http://arxiv.org/abs/2206.06313}{{\ttfamily arXiv:2206.06313
  [astro-ph.CO]}}.

\bibitem{Vegetti:2018dly}
S.~Vegetti, G.~Despali, M.~R. Lovell, and W.~Enzi, ``{Constraining sterile
  neutrino cosmologies with strong gravitational lensing observations at
  redshift z \ensuremath{\sim} 0.2},''
  \href{http://dx.doi.org/10.1093/mnras/sty2393}{{\em Mon. Not. Roy. Astron.
  Soc.} {\bfseries 481} no.~3, (2018) 3661--3669},
  \href{http://arxiv.org/abs/1801.01505}{{\ttfamily arXiv:1801.01505
  [astro-ph.CO]}}.

\bibitem{Gilman:2019vca}
D.~Gilman, S.~Birrer, T.~Treu, A.~Nierenberg, and A.~Benson, ``{Probing dark
  matter structure down to $10^7$ solar masses: flux ratio statistics in
  gravitational lenses with line-of-sight haloes},''
  \href{http://dx.doi.org/10.1093/mnras/stz1593}{{\em Mon. Not. Roy. Astron.
  Soc.} {\bfseries 487} no.~4, (2019) 5721--5738},
  \href{http://arxiv.org/abs/1901.11031}{{\ttfamily arXiv:1901.11031
  [astro-ph.CO]}}.

\bibitem{Gilman:2019nap}
D.~Gilman, S.~Birrer, A.~Nierenberg, T.~Treu, X.~Du, and A.~Benson, ``{Warm
  dark matter chills out: constraints on the halo mass function and the
  free-streaming length of dark matter with eight quadruple-image strong
  gravitational lenses},'' \href{http://dx.doi.org/10.1093/mnras/stz3480}{{\em
  Mon. Not. Roy. Astron. Soc.} {\bfseries 491} no.~4, (2020) 6077--6101},
  \href{http://arxiv.org/abs/1908.06983}{{\ttfamily arXiv:1908.06983
  [astro-ph.CO]}}.

\bibitem{Nadler:2021dft}
E.~O. Nadler, S.~Birrer, D.~Gilman, R.~H. Wechsler, X.~Du, A.~Benson, A.~M.
  Nierenberg, and T.~Treu, ``{Dark Matter Constraints from a Unified Analysis
  of Strong Gravitational Lenses and Milky Way Satellite Galaxies},''
  \href{http://dx.doi.org/10.3847/1538-4357/abf9a3}{{\em Astrophys. J.}
  {\bfseries 917} no.~1, (2021) 7},
  \href{http://arxiv.org/abs/2101.07810}{{\ttfamily arXiv:2101.07810
  [astro-ph.CO]}}.

\bibitem{Montel:2022fhv}
N.~A. Montel, A.~Coogan, C.~Correa, K.~Karchev, and C.~Weniger, ``{Estimating
  the warm dark matter mass from strong lensing images with truncated marginal
  neural ratio estimation},'' \href{http://arxiv.org/abs/2205.09126}{{\ttfamily
  arXiv:2205.09126 [astro-ph.CO]}}.

\bibitem{Lee:2020wfn}
V.~S.~H. Lee, A.~Mitridate, T.~Trickle, and K.~M. Zurek, ``{Probing Small-Scale
  Power Spectra with Pulsar Timing Arrays},''
  \href{http://dx.doi.org/10.1007/JHEP06(2021)028}{{\em JHEP} {\bfseries 06}
  (2021) 028}, \href{http://arxiv.org/abs/2012.09857}{{\ttfamily
  arXiv:2012.09857 [astro-ph.CO]}}.

\bibitem{Lee:2021zqw}
V.~S.~H. Lee, S.~R. Taylor, T.~Trickle, and K.~M. Zurek, ``{Bayesian Forecasts
  for Dark Matter Substructure Searches with Mock Pulsar Timing Data},''
  \href{http://dx.doi.org/10.1088/1475-7516/2021/08/025}{{\em JCAP} {\bfseries
  08} (2021) 025}, \href{http://arxiv.org/abs/2104.05717}{{\ttfamily
  arXiv:2104.05717 [astro-ph.CO]}}.

\bibitem{Kashiyama:2018gsh}
K.~Kashiyama and M.~Oguri, ``{Detectability of Small-Scale Dark Matter Clumps
  with Pulsar Timing Arrays},''
  \href{http://arxiv.org/abs/1801.07847}{{\ttfamily arXiv:1801.07847
  [astro-ph.CO]}}.

\bibitem{Delos:2021rqs}
M.~S. Delos and T.~Linden, ``{Dark Matter Microhalos in the Solar Neighborhood:
  Pulsar Timing Signatures of Early Matter Domination},''
  \href{http://arxiv.org/abs/2109.03240}{{\ttfamily arXiv:2109.03240
  [astro-ph.CO]}}.

\bibitem{Clark:2015sha}
H.~A. Clark, G.~F. Lewis, and P.~Scott, ``{Investigating dark matter
  substructure with pulsar timing -I. Constraints on ultracompact
  minihaloes},'' \href{http://dx.doi.org/10.1093/mnras/stv2743}{{\em Mon. Not.
  Roy. Astron. Soc.} {\bfseries 456} no.~2, (2016) 1394--1401},
  \href{http://arxiv.org/abs/1509.02938}{{\ttfamily arXiv:1509.02938
  [astro-ph.CO]}}. [Erratum: Mon.Not.Roy.Astron.Soc. 464, 2468 (2017)].

\bibitem{Ishiyama:2010es}
T.~Ishiyama, J.~Makino, and T.~Ebisuzaki, ``{Gamma-ray Signal from Earth-mass
  Dark Matter Microhalos},''
  \href{http://dx.doi.org/10.1088/2041-8205/723/2/L195}{{\em Astrophys. J.
  Lett.} {\bfseries 723} (2010) L195},
  \href{http://arxiv.org/abs/1006.3392}{{\ttfamily arXiv:1006.3392
  [astro-ph.CO]}}.

\bibitem{Bond:1990iw}
J.~R. Bond, S.~Cole, G.~Efstathiou, and N.~Kaiser, ``{Excursion set mass
  functions for hierarchical Gaussian fluctuations},''
  \href{http://dx.doi.org/10.1086/170520}{{\em Astrophys. J.} {\bfseries 379}
  (1991) 440}.

\bibitem{Bower:1991kf}
R.~G. Bower, ``{The Evolution of groups of galaxies in the Press-Schechter
  formalism},'' {\em Mon. Not. Roy. Astron. Soc.} {\bfseries 248} (1991) 332.

\bibitem{Hiroshima:2022khy}
N.~Hiroshima, S.~Ando, and T.~Ishiyama, ``{Semi-analytical frameworks for
  subhalos from the smallest to the largest scale},''
  \href{http://arxiv.org/abs/2206.01358}{{\ttfamily arXiv:2206.01358
  [astro-ph.CO]}}.

\bibitem{Bland-Hawthorn:2016}
J.~Bland-Hawthorn and O.~Gerhard, ``{The Galaxy in Context: Structural,
  Kinematic, and Integrated Properties},''
  \href{http://dx.doi.org/10.1146/annurev-astro-081915-023441}{{\em Ann. Rev.
  Astron. Astrophys.} {\bfseries 54} (2016) 529},
  \href{http://arxiv.org/abs/1602.07702}{{\ttfamily arXiv:1602.07702
  [astro-ph.GA]}}.

\bibitem{Correa:2015dva}
C.~A. Correa, J.~S.~B. Wyithe, J.~Schaye, and A.~R. Duffy, ``{The accretion
  history of dark matter haloes \textendash{} III. A physical model for the
  concentration\textendash{}mass relation},''
  \href{http://dx.doi.org/10.1093/mnras/stv1363}{{\em Mon. Not. Roy. Astron.
  Soc.} {\bfseries 452} no.~2, (2015) 1217--1232},
  \href{http://arxiv.org/abs/1502.00391}{{\ttfamily arXiv:1502.00391
  [astro-ph.CO]}}.

\bibitem{Hu:2002we}
W.~Hu and A.~V. Kravtsov, ``{Sample variance considerations for cluster
  surveys},'' \href{http://dx.doi.org/10.1086/345846}{{\em Astrophys. J.}
  {\bfseries 584} (2003) 702--715},
  \href{http://arxiv.org/abs/astro-ph/0203169}{{\ttfamily
  arXiv:astro-ph/0203169}}.

\bibitem{Ishiyama:2011af}
T.~Ishiyama, J.~Makino, S.~Portegies~Zwart, D.~Groen, K.~Nitadori, S.~Rieder,
  C.~de~Laat, S.~McMillan, K.~Hiraki, and S.~Harfst, ``{The Cosmogrid
  Simulation: Statistical Properties of Small Dark Matter Halos},''
  \href{http://dx.doi.org/10.1088/0004-637X/767/2/146}{{\em Astrophys. J.}
  {\bfseries 767} (2013) 146}, \href{http://arxiv.org/abs/1101.2020}{{\ttfamily
  arXiv:1101.2020 [astro-ph.CO]}}.

\bibitem{Bryan:1997dn}
G.~L. Bryan and M.~L. Norman, ``{Statistical properties of x-ray clusters:
  Analytic and numerical comparisons},''
  \href{http://dx.doi.org/10.1086/305262}{{\em Astrophys. J.} {\bfseries 495}
  (1998) 80}, \href{http://arxiv.org/abs/astro-ph/9710107}{{\ttfamily
  arXiv:astro-ph/9710107}}.

\bibitem{Penarrubia_2010}
J.~Penarrubia, A.~J. Benson, M.~G. Walker, G.~Gilmore, A.~W. McConnachie, and
  L.~Mayer, ``The impact of dark matter cusps and cores on the satellite galaxy
  population around spiral galaxies,''
  \href{http://dx.doi.org/10.1111/j.1365-2966.2010.16762.x}{{\em Monthly
  Notices of the Royal Astronomical Society} (May, 2010) no--no}.
  \url{https://doi.org/10.1111%2Fj.1365-2966.2010.16762.x}.

\bibitem{Lewis:1999bs}
A.~Lewis, A.~Challinor, and A.~Lasenby, ``{Efficient computation of CMB
  anisotropies in closed FRW models},''
  \href{http://dx.doi.org/10.1086/309179}{{\em Astrophys. J.} {\bfseries 538}
  (2000) 473--476}, \href{http://arxiv.org/abs/astro-ph/9911177}{{\ttfamily
  arXiv:astro-ph/9911177}}.

\end{thebibliography}\endgroup

\end{document}